\documentclass[a4paper,onecolumn,11pt,accepted=2023-01-13]{quantumarticle}
  \pdfoutput=1
\usepackage{fullpage}
\usepackage{amsmath, amssymb, amsthm, amsfonts, latexsym, color, paralist,url,ifdraft,lmodern,wrapfig}
\usepackage{thm-restate}
\usepackage{csquotes}
\usepackage[T1]{fontenc}
\usepackage[final]{graphicx}
\usepackage{silence} 
\usepackage[refpage,prefix]{nomencl}
\usepackage[draft=false,pageanchor]{hyperref}

\usepackage{algorithm,algorithmicx,algpseudocode}
\newcommand*{\nom}[2]{#1\nomenclature{#1}{#2}}
\usepackage[displaymath,textmath,sections,graphics,floats]{preview}
\makenomenclature
\usepackage{bbold}
\usepackage{pdflscape,geometry} 
\usepackage{makecell} 

  \theoremstyle{plain}
  \newtheorem{theorem}{Theorem}
  \newtheorem{lemma}[theorem]{Lemma}
  \newtheorem{corollary}[theorem]{Corollary}

  \theoremstyle{definition}
  \newtheorem{definition}[theorem]{Definition}

  \theoremstyle{remark}
  \newtheorem{remark}{Remark}
  \theoremstyle{plain}
  \newtheorem*{theorem*}{Theorem}
  \newtheorem*{lemma*}{Lemma}
  \newtheorem*{corollary*}{Corollary}
  \newtheorem*{proposition*}{Proposition}
  \newtheorem*{claim*}{Claim}
  
\ifdraft{\newcommand{\authnote}[3]{{\color{#3} {\bf  #1:} #2}}}{\newcommand{\authnote}[3]{}}

\newcommand{\onote}[1]{\authnote{Or}{#1}{green}}

\newcommand{\poly}{\mathrm{poly}}

\newcommand{\bra}[1]{\langle #1 \vert}

\newcommand{\ket}[1]{\vert #1 \rangle}
\DeclareMathOperator{\tr}{Tr}

\newcommand{\NN}{\mathbb{N}}

\newcommand{\ketbra}[1]{\vert #1 \rangle \langle #1 \vert}
\newcommand{\braket}[2]{\langle #1 \vert #2 \rangle}

\newcommand{\tensor}{\otimes}

\DeclareMathOperator{\E}{\mathbb{E}}
\PreviewMacro*\onote
\PreviewMacro*\nom
\PreviewMacro*\ifdraft
\DeclareRobustCommand{\rchi}{{\mathpalette\irchi\relax}}
\newcommand{\irchi}[2]{\raisebox{\depth}{$#1\chi$}}
\newcommand{\getsr}{\stackrel {R}{\gets}}
\newcommand{\PA}{\mathbb{P}_A}
\newcommand{\PAD}{\mathbb{P}_{A^\perp}}
\newcommand{\io}{\ensuremath{i\mathcal{O}}}

\newcommand{\keygen}{\ensuremath{\mathsf{key\textit{-}gen}}}
\newcommand{\tokengen}{\ensuremath{\mathsf{token\textit{-}gen}}}
\newcommand{\sign}{\ensuremath{\mathsf{sign}}}
\newcommand{\verify}{\ensuremath{\mathsf{verify}}}
\newcommand{\hindex}{\ensuremath{\mathsf{index}}}
\newcommand{\decrypt}{\ensuremath{\mathsf{decrypt}}}
\newcommand{\encrypt}{\ensuremath{\mathsf{encrypt}}}

\newcommand{\revoke}{\ensuremath{\mathsf{revoke}}}
\newcommand{\verifytoken}{\ensuremath{\mathsf{verify\textit{-}token}}}
\newcommand{\cnt}{\ensuremath{\mathsf{count}}} 

\newcommand{\otr}{\ensuremath{\mathsf{OTR}}}
\newcommand{\ot}{\ensuremath{\mathsf{OT}}}
\newcommand{\ts}{\ensuremath{\mathsf{TS}}}
\newcommand{\ds}{\ensuremath{\mathsf{DS}}}
\newcommand{\tm}{\ensuremath{\mathsf{TM}}}

\newcommand{\negl}{\ensuremath{\mathsf{negl}}}
\newcommand{\nonnegl}{\ensuremath{\mathsf{non\textit{-}negl}}}

\newcommand{\Adv}{\textsf{Adv}}
\newcommand{\Sim}{\textsf{Sim}}
\newcommand{\Dist}{\textsf{Dist}}

\newcommand{\IA}{\ensuremath{\rchi_{A}}}
\newcommand{\IAD}{\ensuremath{\rchi_{A^{\perp}}}}
\newcommand{\IAH}{\ensuremath{\rchi_{A^*}}}

\newcommand{\stamp}{
 {\mathchoice
  {\includegraphics[height=1.8ex]{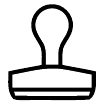}}
  {\includegraphics[height=1.8ex]{RubberStamp}}
  {\includegraphics[height=1.2ex]{RubberStamp}}
  {\includegraphics[height=0.9ex]{RubberStamp}}
 }
}
\usepackage{booktabs}
\usepackage{mathtools}
\mathtoolsset{showonlyrefs}

\begin{document}
\title{Quantum Tokens for Digital Signatures}
\author[1]{Shalev Ben-David}
\affil[1]{University of Waterloo,  David R. Cheriton School of Computer Science}
\author[2]{Or Sattath}
 \orcid{0000-0001-7567-3822}
\affil[2]{Ben-Gurion University of the Negev, Computer Science Department}

\maketitle
\begin{abstract}
The fisherman caught a quantum fish.
\textit{Fisherman, please let me go}, begged the fish,  \textit{and I will grant you three wishes}. The fisherman agreed. The fish gave the fisherman a quantum computer, three quantum signing tokens and his classical public key.
The fish explained: \textit{to sign your three wishes, use the tokenized signature scheme on this quantum computer, then show your valid signature to the king who owes me a favor}.

The fisherman used one of the signing tokens to sign the document ``give me a castle!'' and rushed to the palace. The king executed the classical verification algorithm using the fish's public key, and since it was valid, the king complied. 

The fisherman's wife wanted to sign ten wishes using their two remaining signing tokens. The fisherman did not want to cheat, and secretly sailed to meet the fish. \textit{Fish, my wife wants to sign ten more wishes.} But the fish was not worried: \textit{I have learned quantum cryptography following the previous story}\footnote{\href{http://www.pitt.edu/~dash/grimm019.html}{The Fisherman and His Wife}
by the brothers Grimm.}. \textit{These quantum tokens are consumed during the signing. 
Your polynomial wife cannot even sign four wishes using the three signing tokens I gave you}.

\textit{How does it work?} wondered the fisherman. \textit{Have you heard of quantum money? These are quantum states which can be easily verified but are hard to copy. This tokenized quantum signature scheme extends Aaronson and Christiano's quantum money scheme~\cite{aaronson13quantum},
which is why the signing tokens cannot be copied}.

\textit{Does your scheme have additional fancy properties?} asked the fisherman.
\textit{Yes, the scheme has other security guarantees: revocability, testability and everlasting revocability. Furthermore, if you're at sea and your quantum phone has only classical reception, you can use this scheme to transfer the value of the quantum money to shore}, said the fish, and swam away.
\begin{center}
\textbf{The Normal Abstract}    
\end{center}
 We introduce a new quantum cryptographic primitive which we call
a tokenized signature scheme.
Ordinary digital signatures are used to produce signed documents which are publicly verifiable but are unfeasible to forge by third parties.
A tokenized signature scheme has the additional property that the signer can produce and distribute one-use quantum signing tokens that allow the holder
to sign one (and only one) document of her choice.

We construct a tokenized signature scheme based
on Aaronson and Christiano's quantum money~\cite{aaronson13quantum}, and discuss its security.
We also show how other notions of security could be achieved,
including revocability, testability, and everlasting revocability. 
Finally, we show that any testable tokenized signature scheme can be used as a quantum money scheme with additional desirable properties, including the ability to turn a quantum bill into a classical check.
\end{abstract}
\tableofcontents

\onote{TODO:
- Before submitting to the arXiv, remove onotes, snotes, comments etc. }
\section{Introduction}

One of the main goals of cryptography is to allow an authorized party,
typically holding a secret key, to perform an action which an unauthorized
party (without the key) cannot.
For example, in a digital signature scheme, the authorized party, Alice, holds
a secret key that allows her to create digital signatures that will be accepted
by a public verification algorithm. Anyone without Alice's
key cannot forge her signature. 

In this work,
we consider the task of delegating \emph{limited} authorization:
is it possible to provide a one-time access to the secret key to a third party? For example, if Alice goes on vacation, can she allow Bob to sign one (and only one) document of his choice?

Classically, Bob either knows the secret key or doesn't,
and there is no way to control how many times the key is used. But
with quantum mechanics, the situation is different: the no cloning theorem~\cite{wootters1982single}
allows us to create secrets that cannot be copied. Consequently, we propose the design
of cryptographic schemes with two levels of secrets: one classical 
``master'' secret, which is used only to generate any number of
unclonable quantum ``tokens'', each of which can be used to perform one action, and is consumed in the process due to destructive measurements. If Alice holds the secret key, she can
delegate authorization to Bob by granting him a limited number
of quantum tokens.

\subsection{Tokens for Digital Signatures}
\label{sec:toekns_for_digital_signatures}
This work applies the previous proposal regarding tokens specifically to digital signatures, allowing the delegation of limited authorization via the use of quantum tokens.

Digital signature is a cryptographic primitive which is arguably second in importance only to encryption.
A digital signature scheme~\cite{diffie76new} consists of three Probabilistic Polynomial Time (\nom{PPT}{Probabilistic Polynomial Time}) algorithms: \keygen, \sign, and \verify.
The first algorithm outputs a secret key $sk$ and a public key $pk$. The signer can use $sk$ to sign a document \nom{$\alpha$}{A document or a message, that usually needs to be signed} by calling $\sign(sk,\alpha)$.
This produces a signature $sig$,
which can be verified by anyone holding the public key
by calling $\verify(pk,\alpha,sig)$. We will denote the keys as subscripts,
so these calls are $\sign_{sk}(\alpha)$ and $\verify_{pk}(\alpha,sig)$. The \verify\ algorithm returns either 1 (interpreted as ``accept'')
or 0. 
A valid signature should be accepted, so $\verify_{pk}(\alpha,\sign_{sk}(\alpha))$
should accept for all $\alpha$.
A digital signature scheme is 
\emph{secure against chosen message attack} if a polynomial time adversary with access to a signing oracle cannot efficiently generate any \emph{fresh} signature (that is, a pair $\alpha,sig$ for which $\verify_{pk}(\alpha,sig)$ accepts, where $\alpha$ was not signed by the oracle), except with a negligible probability. 

Our main contribution is a construction of a quantum \emph{tokenized} signature
scheme.
A tokenized signature scheme consists of 4 quantum polynomial time (\nom{QPT}{Quantum Polynomial Time} for short) algorithms: \keygen,\ \tokengen,\ \sign,\ and \verify.
In this setting there are \emph{three} entities: a signer, a verifier and a new entity, which we will call the signing authority.  In this context, even though the signer is the one signing the document, it is done in the name of the signing authority.
The authority generates the pair $(sk,pk)$ using $\keygen$ as before. Next, it generates a quantum state \nom{$\ket{\stamp}$}{A quantum signing token}, which we call a \emph{signing token}, by running $\tokengen_{sk}$.\footnote{The symbol $\stamp$ represents a rubber stamp.} We emphasize that different calls to $\tokengen_{sk}$ may produce different signing tokens. The signer, who gets one copy of a signing token from the authority,
can sign a single document of her choice. The output of $\sign(\alpha,\ket{\stamp})$ is classical, similarly to the classical setting. The correctness property is very similar:
$\verify_{pk}(\alpha,\sign(\alpha,\ket{\stamp}))$ must accept for all
documents $\alpha$.

The novelty of tokenized signatures is that \sign\ applies a measurement which collapses and hence \emph{consumes} $\ket{\stamp}$, and therefore it cannot be reused to sign an additional document. Informally, the security requirement is that a QPT adversary \nom{$\Adv$}{A QPT adversary}, with access to the public key $pk$ and to $\ell$\nomenclature{$\ell$}{The number of signing tokens given to the adversary} signing tokens $\ket{\stamp_{1}}\tensor \ldots \tensor \ket{\stamp_{\ell}}$
,
cannot generate valid signatures for $\ell+1$ different documents.

Here are two motivating examples.
\begin{itemize}
    \item  A manager wants to hedge the embezzlement risks that an accountant introduces. This can be achieved by agreeing with the bank that any signed wire is limited to \$1000. Then, the manager can grant the accountant a number of tokens according to her level of trust in the accountant.
    \item Online computers are more prone to hacks than offline computers. A system administrator can hold the secret keys on an offline computer, and generate signing tokens to be used on the online computer.
\end{itemize}
The common denominator of these examples is that the usage of tokens reduces risk, as the potential damage of abusing a token is smaller than that of the secret key. 

A Tokenized signature scheme can be used as a digital signature scheme, up to some minor technicalities (see Theorem~\ref{thm:ts_implies_mds}). 
Every tokenized digital scheme is also \emph{revocable}:
the signer can destroy a token in a publicly-verifiable way.
A motivating example is an employee who receives signing tokens which were generated by the system administrator. Once the employee resigns, the employer can ask her to revoke her tokens. The employer only needs the public key (not the secret key, perhaps kept only by the system administrator) to run $\revoke$.\footnote{The employee can provide the signed documents for the signing tokens that were already used, and the employer can determine whether they are valid and meet the firm's policies.}
If the (proclaimed) signing token passes revocation it is guaranteed that no unauthorized document was or will be signed by the employee.\footnote{For this to work, the employee must cooperate and provide the previously signed documents and remaining signing tokens. This mechanism is not suitable for workers that do not cooperate, or other scenarios such as an attacker who gains access to the signing tokens.} 
The algorithm $\revoke$ is extremely simple (see Section \ref{sec:properties-tokenized-signatures}): use the (proclaimed) signing token to sign a random document, and verify the signature's validity. The formal statement is given in Theorem~\ref{thm:revocability}. 

An even stronger notion of revocability is testability. In this case,
the algorithm $\verifytoken$ allows testing the signing token without consuming it.
Not every tokenized signature scheme is necessarily testable, but the scheme we construct is. We show that every testable tokenized signature
scheme can also be used as a public quantum money scheme
(see Theorem~\ref{thm:testable_ts_is_qm}).

Another interesting notion which can some of our schemes achieve is everlasting revocability. A protocol has everlasting revocability if it is secure against computationally unbounded adversaries \emph{after} it ends --- see \cite{DR02,Unr15,Unr18} and references therein. 
In our context it means that a computationally unbounded quantum adversary $\Adv$ with access to a single copy of $\ket{\stamp}$ but without access to $pk$, cannot pass revocation, and then, perhaps decades later, also generate a valid signature for some
document $\alpha$.
Here is a motivating example. Consider a queen who is going on a mission. The queen provides the signing token to the heir to be used in case of an emergency,  and the public key to the lawyer, accountant, and family doctor (but \emph{not} the heir). Recall that the public key (which the heir does not have access to) is not needed for signing -- it is only required for the verification. When the queen returns from her mission, she may ask the heir to return the signing token. If $\revoke$ passes, it is guaranteed that even a computationally unbounded heir cannot generate a valid signature (except a negligible probability). 

\subsection{The Construction}
We first reduce the problem of constructing a tokenized
digital signature scheme to the problem of
constructing a \emph{single-bit} tokenized
signature scheme, which can only sign a single document consisting
of a single bit. That is, we show how to extend any single-bit one-time scheme into a full blown tokenized digital signature scheme.

This extension is done in three steps.
The first step is to extend the scheme
to support a single document consisting of $r$ bits. This can be done
by essentially duplicating the single-bit protocol $r$ times. In the second step, we extend the scheme to support
documents of arbitrary length --- rather than a length $r$ which is fixed in advance ---
by employing a hash function (which we require to be collision-resistant
against a quantum adversary). The idea is to hash the document
down to $r$ bits, and then to sign the hash. This is standard technique known as the hash-and-sign paradigm.

At this point, the scheme can sign a single document using a single
token. Next, we need to extend it to a scheme that can support
an arbitrary number of tokens. We do this by
employing a classical digital signature scheme, which we require to be
secure against chosen message attacks by a quantum adversary.
The public key of the final tokenized scheme will be the public key
of the classical signature scheme. Then, every time a token is required,
it is generated by restarting the one-time tokenized scheme from scratch
(using a new public and secret key each time), signing the public
key, using the classical signature scheme, and attaching it to the token.

In Section~\ref{sec:standard-construction}, we prove the security
of these three steps, and also show that they preserve
several desirable properties. This reduces the problem of constructing a tokenized signature scheme to the problem of constructing a single-bit, one-time tokenized signature scheme.

The main challenge is to construct the single-bit tokenized scheme.
Our construction for this is based on Aaronson and Christiano's quantum money scheme~\cite{aaronson2012quantum,aaronson13quantum}. In general, we do not know
how to use a public quantum money scheme to provide a tokenized
signature scheme in a black-box way; however, we can use the specific
construction of \cite{aaronson13quantum}, based on hidden subspaces,
to construct a tokenized signature scheme with many desirable properties.

Informally, the construction of Aaronson and Christiano works as follows.
They consider a random subspace $A$ of $\mathbb{F}^n_2$ of dimension $n/2$,
and the money state (in our case, this would be the signing token) uses $n$ qubits,  which are in the uniform superposition over $A$,
denoted by $\ket{A}$.
For example, for $n=4$, the space $\mathbb{F}_2^4$ consists of 16 vectors $0000,\, 0001,\ldots,\, 1111$, and one choice for the subspace $A$ could be the vectors $0000,\,0011,\,1110,\, 1101$. The addition operation is bitwise addition modulo 2 (XOR), for example: $0011\oplus 1110 = 1101$. The state $\ket{A}$ in this case is the 4 qubit state (a unit vector in $\mathbb{C}^{2^4}$): $\frac{1}{2} (\ket{0000}+\ket{0011}+\ket{1110}+\ket{1101})$.

Another important subspace of $\mathbb{F}_2^n$ is $A^\perp$: 
\[A^\perp=\{b\in \mathbb{F}_2^n| \forall a \in A,\ a\cdot b \equiv \sum_{i=1}^n a_i b_i \mod 2 =0   \}.\]
In the example above, $A^\perp$ consists of the vectors $0000,\, 0111,\, 1011,\, 1100$. Applying  $H^{\otimes n}$ (which can be implemented efficiently on a quantum computer by a depth 1 circuit) 
on $\ket{A}$ gives the state $\ket{A^\perp}$: 
\[ H^{\otimes n}\ket{A}=\ket{A^\perp}=\frac{1}{2^{n/4}}\sum_{b \in A^\perp} \ket{b},\]  
where $H=\frac{1}{\sqrt{2}}\begin{pmatrix}1 & 1 \\ 1 & -1\end{pmatrix} $.

We can use the above properties to construct a \emph{private} tokenized signature scheme. In the classical literature, there are two types of digital signatures: the standard one, which was described above; the other is a private (symmetric) digital signature scheme, known as a \emph{message authentication code} (MAC), in which verification requires the secret key. The difference between (public) digital signature scheme and (symmetric) MAC is very similar to the difference between public and symmetric key encryption.\footnote{An alternative term would be tokenized MAC, but we use private tokenized signature scheme, to follow the same conventions that were made between public and private, quantum money.}

The signing authority samples a random $n/2$ dimensional subspace $A$ as the secret key, and generates $\ket{A}$ as the signing token. The idea is to let any non-zero vector in $A$ correspond to a signature for the bit $0$, and a non-zero
vector in $A^\perp$ correspond to a signature
for the bit $1$. An adversary holding $\ket{A}$ can either measure it in the standard basis to get a (uniformly random) element of $A$, or can measure
$\ket{A^\perp}$ to get an element of $A^\perp$. We show that an adversary with a single copy of $\ket{A}$ cannot do both, since measuring the state collapses it. This is a manifestation of the no-cloning theorem~\cite{wootters1982single}: if the adversary could clone the state $\ket{A}$, and hold two copies of it, she  could use the first copy to find an element of $A$, and the second copy to find an element of $A^\perp$, and break the security of the scheme.
The verifier, which in the private setting knows the secret key (the choice of $A$ and $A^\perp$), can verify whether the signature is valid, by testing for the relevant subspace membership. 
This gives us a private tokenized signature scheme whose
only security assumption is the existence of a collision-resistant hash function secure against quantum adversaries, with all the analogous properties of the public tokenized signature scheme --- see Appendix~\ref{sec:MAC}. 

How can we turn this into a public tokenized signature scheme? Suppose the adversary has one copy of $\ket{A}$ as before, but additionally can ask questions of the following form in superposition: is $x$ in $A$? Is $y$ in $A^\perp$? 
We use quantum query complexity techniques to prove that such an adversary still cannot efficiently find both a non-zero element of $A$ and of $A^\perp$ (see Section~\ref{sec:oracle-results}). 
These questions are sufficient to verify signatures. Therefore, the problem we face is the following: is there a way to obfuscate the membership for $A$ and $A^\perp$? The obfuscated program should allow the verifier to test whether an element is in the relevant subspace, but should not leak any additional information about $A$ and $A^\perp$. 
Aaronson and Christiano faced the same challenge and suggested an ad-hoc approach for obfuscating these subspaces. 
Unfortunately, a variant of their scheme, known as the noise-free case, was broken in~\cite{Pena15Algebraic}. This attack can be extended to the (main) noisy scheme, so that it is also broken; this observation
was made by Paul Christiano, and is reported here in Section~\ref{sec:attack-noisy-scheme} for completeness. Interestingly, Christiano's attack is inherently quantum, and a classical attack was shown subsequently~\cite{PDF+19}.

Earlier versions of this work conjectured that a scheme in which the public key is the obfuscated membership programs for $A$ and $A^\perp$ would be secure. Subsequently to this work, Zhandry introduced a new method to recover from the attacks mentioned above on quantum money, using indistinguishability obfuscation ($\io$)~\cite{zhandry17quantum}.


See Section~\ref{sec:attack_fix_quantum_money} for discussion.

\subsection{Computational Requirements}
One of the functions of money is store of value. From an engineering perspective, this makes quantum money (which our construction is based on) a challenging task, as it requires a long term quantum memory to store the quantum money, which is beyond the capabilities of the current state of the art technology (see, for example, ~\cite{zhong2015optically}). Signing tokens do not necessarily need to be stored for long periods (perhaps, only the signed documents are), and therefore some applications may require relatively short term memory, and will thus be easier to implement in practice.
Furthermore, signing requires a very simple (depth $1$) quantum circuit, and verification is done on a classical computer.
Our tokens can be generated using Clifford circuits, which do not require a universal quantum computer, and therefore may be easier to implement. The only task which requires a universal quantum computer is $\verifytoken$ -- which is not part of the minimal requirements for tokenized signatures, but is required for constructing public quantum money out of it. 
The security of the weak scheme is exponential in the number of qubits;
this means, among other things, that the depth of $\tokengen$ in all our schemes is roughly logarithmic in the security parameter.

\subsection{Application to Quantum Money}

We show that a testable tokenized signature scheme can be used as quantum money\cite{wiesner1983conjugate,bennett1983quantum,tokunaga2003anonymous,mosca2010quantum,aaronson2009quantum,gavinsky2012quantum,aaronson13quantum,farhi2012quantum,georgiou15new}.
Interestingly, the quantum money scheme we get this way has
a combination of desirable properties that no previous scheme had,
not even the Aaronson-Christiano scheme on which our construction
is based.

To get a public quantum money scheme from a public testable tokenized signature scheme, simply use the signing tokens as money. We show in Theorem~\ref{thm:testable_ts_is_qm} that this satisfies the definition
of a public quantum money scheme. An interesting property of this scheme
is that the money can be converted into a publicly-verifiable signature
for a document. This can be used to send the money over a classical channel (vis-\`a-vis ``standard" quantum money, which can only be sent via a quantum channel).

In particular, suppose Alice wants to send money to Bob,
but she is stranded and cannot get quantum internet reception on her quantum cellphone. 
Can she send the money using only classical communication? Some private quantum money schemes~\cite{gavinsky2012quantum,Pastawski02102012} are ``classically verifiable'', meaning, there is a two-way communication protocol between the bank and the user which is used for verification. 
With our scheme, Alice can use her money as a signing token to sign the document
``I'm sending this money, with the serial number 03217, to Bob''.
Bob can then take the signed document and present it to his bank.
The bank knows that Alice must have burned her money to produce
the signed document, and so it can safely issue Bob a new money state.
In essence, Alice can convert a quantum bill into a classical ``check''. This technique, as well as it advantages (it requires only one-way communication) and disadvantages (bank branches need to maintain a database) are discussed in Section~\ref{sec:mult-bank-branch}.


This alternate payment method is also secure against a variety of attacks.
For instance, anyone eavesdropping on the transaction cannot get access
to the money (indeed, Alice may as well broadcast her payment publicly). On the other hand an adversary controlling the quantum communication channel between the sender and the receiver could destroy the quantum money (or even steal it if the channel is not encrypted). 
Furthermore, the bank branch needs only the public key in order to verify
that the payment is valid, meaning that individual branches do not
need to store the secret key at all -- so if the branch is hacked, it loses the quantum money it stored in its safe, but no further damage to the monetary system is inflicted; the thief does not gain the ability to mint new money.

\subsection{Security Assumptions}
\label{sec:security_assuptions}

In light of the failure of various recent cryptographic schemes,
we should carefully discuss our security assumptions. 

The mentioned constructions that were broken lacked a security proof, or used a specific, non-standard assumption. Our work uses only general assumptions.

Our construction suffices for a private tokenized signature scheme, which we refer to as quantum tokens for MAC (see Appendix~\ref{sec:MAC}), under very mild cryptographic assumptions: the existence of a
collision-resistant hash function (secure against quantum adversaries).
There are various candidates for collision-resistant hash functions
that are believed to be secure, so this construction appears to be
on solid ground. All of our main results apply equally well for the private scheme---see Appendix~\ref{sec:MAC}.

Several of our results,
such as the construction of public quantum
money (that can be turned into classical ``checks'') from any public tokenized signature scheme, as well as many others (Theorems~\ref{thm:revocability},~\ref{thm:ts_implies_mds},~\ref{thm:standard_construction_unforgability},~\ref{thm:standard_construction_testable},~\ref{thm:standard_construction_everlasting}) hold without any assumptions.
For this reason, we believe that a public tokenized signature scheme is an interesting primitive even if our current construction fails.


For the construction of a \emph{public} tokenized digital signature scheme, we can show
our protocol is provably secure relative to an oracle. Instantiating
the oracle can be done via a certain type of black-box obfuscation
for a specific class of functions\footnote{Recall that while general black-box obfuscation is impossible~\cite{barak12on}, the impossibility result does not rule out black-box obfuscations for restricted classes of functions.};
we hence prove a reduction from
tokenized digital signatures to a certain type of obfuscation.
The obfuscation part is the only problematic step.
There has been a virtual black-box obfuscation proposal for a class of function that are similar to what we need, but it does not 
exactly match our requirements (see Section~\ref{sec:one-time_scheme}). 
In a previous version of this manuscript, we conjectured that we could replace the VBB assumption with post-quantum indistinguishability obfuscation ($\io$). This turned out to be (almost) correct: for more details see Section~\ref{sec:subsequent_works}. Recently, a pre-quantum $\io$ was constructed based on standard assumptions. Yet, there are only candidate constructions for post-quantum $\io$ --- for more details, see~\cite{CLLZ21}. 


\subsection{Subsequent Works}
\label{sec:subsequent_works}
Since this work first appeared, the field, which is now often called uncloneable cryptography, has been developed in various directions; For a recent review, see~\cite{Sat22b}.

This work presents a construction for tokenized signature relative to a classical oracle (see Section~\ref{sec:oracle-results}), and another one that is based on virtual-black-box (VBB) obfuscation (see Section~\ref{sec:one-time_scheme}). Unfortunately, there are no constructions for such VBB obfuscation for subspaces. We conjectured that replacing the VBB obfuscation with post-quantum indistinguishability obfuscation, which is a heuristic that works in many cases, would work. A variation of our conjecture, which is based on \emph{affine} subspaces, was recently proven by Coladangelo et al.~\cite{CLLZ21}. Whether this variation is needed is still not settled (though we are not aware of any importance in resolving this question). 

Tokens for message authentication codes (MAC)---the symmetric variant of tokens for digiaal signatures (see Appendix~\ref{sec:MAC})---were used to construct disposable back-doors~\cite{CGL19}. Ref.~\cite{BSS21} shows a construction for tokenized MAC which is noise-tolerant and uses only BB84 states. Tokenized signatures were also used to construct uncloneable decryptors~\cite{CLLZ21}. 

Recently, Shmueli~\cite{Shm22} proved a variant of tokenized signatures which is called semi-quantum tokenized signatures. 
Semi-quantum money is a quantum money scheme in which the minting is done in an interactive protocol between the user and a \emph{classical} bank~\cite{RS22,Shm22b}. In semi-quantum tokenized signatures, the generation of the token can be done via an interactive protocol with a classical delegation (in the spirit of the fable of the abstract, the fish does not need to have a quantum computer; only the fisherman needs one).

In Ref.~\cite{AGK+20}, Amos et al. introduced a strengthened version of tokenized signatures called \emph{one-shot} signatures. The main difference in their setting is that there are no secret keys, and no one can generate two valid signatures which are associated with the same public-key. Their construction uses a common reference string (CRS), and their construction is relative to a classical oracle. The question of whether a construction in the standard model exists is open. 

In~\cite{Sat22}, a quantum money scheme with \emph{smart-contracts} capabilities was introduced. The construction is modular in the sense that it can use any tokenized signature, semi-quantum tokenized signature or one-shot signature scheme.
A unified framework that addresses all these different primitives was introduced. In order to do that, the syntax of the tokenized signature scheme was modified---see there for details.


Ref.~\cite{BBK22} studies \emph{constructive post-quantum reductions}. One of their main results shows that certain reductions that hold in the classical setting do not hold in the quantum setting. This is shown using tokenized signatures, using the following approach: classically, one can reduce the problem of signing two random messages to signing one random message by rewinding. Note that this holds even if the adversary is given some auxiliary input. In the quantum setting, this is not the case: given one quantum signing token, it is possible to sign a single random message but not two random messages.  
\subsection{Organization}
\label{sec:structure}
In Section~\ref{sec:preliminaries} we provide preliminary definitions and notation, in Section~\ref{sec:definitions} we introduce new definitions, in Section~\ref{sec:properties-tokenized-signatures} we shows several properties of tokenized signatures schemes. Section~\ref{sec:standard-construction} strengthens a very weak form of tokenized signatures to a full blown tokenized signature scheme. We show a query complexity lower bound in Section~\ref{sec:oracle-results}, which is used later to construct the weak form of tokenized signatures.
In Section~\ref{sec:attack_fix_quantum_money} we revisit Aaronson and Christiano's quantum money scheme, report on an attack by Christiano which makes their original proposal insecure, and then propose how to fix it. This section also contains the weak form of tokenized signature protocol, and we discuss the security  of it. In Section~\ref{sec:applications} we show two applications, where we show how tokenized signatures can be used to (a) transfer the value of quantum money, without using quantum communication, and (b) to rule out \emph{two-faced behavior} in the context of distributed algorithms. 

Nomenclature is provided in Appendix~\ref{sec:nomenclature}. In Appendix~\ref{sec:MAC} we show how to construct a \emph{private} tokenized  signatures scheme, under weaker assumptions than required for the public scheme. Illustration of the abstract can be found in the arXiv auxiliary files available at \url{https://arxiv.org/src/1609.09047/anc}.

\section{Preliminaries and Notation}
\label{sec:preliminaries}

\subsection{Linear Algebra}

For any subspace $A \prec \mathbb{F}_2^n$, let 
\nomenclature{$A^\perp$}{The orthogonal complement of $A$, with respect to the standard dot product in $\mathbb{F}_2^n$}
\[A^\perp=\{b\in \mathbb{F}_2^n| \forall a \in A,\ a\cdot b = \sum_{i=1}^n a_i b_i \mod 2 =0   \}.\]
A basis for $A$ will be denoted \nom{$\langle A \rangle$}{A basis for $A$}.
For any subspace \nom{$A$}{A subspace of $\mathbb{F}_2^n$, usually of dimension $\frac{n}{2}$}, $A^\perp$ is also a subspace, which satisfies 
\begin{equation}
\label{eq:6}
  (A^{\perp})^{\perp}=A
\end{equation}
 and 
 \begin{equation}
   \label{eq:dim_dual_subspaces}
\dim(A)+\dim(A^\perp)=n.
 \end{equation}

We denote by \nom{$\IA$}{The membership function to $A$} the indicator function for $A$, and similarly for $\rchi_{A^\perp}$:
\begin{equation}
  \label{eq:4}   
\IA(a) = 
     \begin{cases}
       1 &\quad a \in A\\
       0 &\quad\text{otherwise.} \\ 
     \end{cases}
\end{equation}
We denote the unified membership function \nom{$\rchi_{A^*}$}{The unified membership to $A$ and $A^\perp$}$:\{0,1\}\times \{0,1\}^n \to \{0,1\}$, which combines $\rchi_A$ and $\rchi_{A^\perp}$:
\begin{equation}
  \label{eq:IAH}   
\IAH(p,a) = 
     \begin{cases}
      \IA(a)&\quad p=0 \\
       \IAD(a) &\quad p=1 \\ 
     \end{cases}
\end{equation}

For even $n$, let \nom{$S(n)$}{The set of all subspaces of $\mathbb{F}_{2}^{n}$ with dimension $\frac{n}{2}$} be the set of all subspaces of $\mathbb{F}_{2}^{n}$ with dimension $\frac{n}{2}$. When $n$ is clear from the context we simply write $S$. For any $A\in S(n)$, we define the state \nom{$\ket{A}$}{The uniform superposition over elements of $A$}: 
\begin{equation}
\ket{A}=\frac{1}{\sqrt{2^{n/2}}} \sum_{a \in A}\ket{a},
\label{eq:ketA}
\end{equation}
and similarly for \nom{$\ket{A^{\perp}}$}{The uniform superposition over elements of $A^\perp$} (here we use Eq.~\eqref{eq:dim_dual_subspaces} to justify the normalization coefficient):
\begin{equation}
\ket{A^{\perp}}=\frac{1}{\sqrt{2^{n/2}}} \sum_{a \in A^{\perp}}\ket{a}.
\label{eq:ketAperp}
\end{equation}

The process in which $x$ is sampled uniformly from a finite set $X$ is denoted by $x\getsr X$, and $output \gets \mathsf{Alg}(input)$ denotes sampling $output$ according to the distribution generated by running $\mathsf{Alg}(input)$.

\subsection{Cryptography}
For a function $f:\mathbb N \to \mathbb{R}^{+}$ we use the shorthand $f(\kappa) \leq \negl(\kappa)$ to say that $f$ is negligible: for every polynomial $p(\kappa)$, there exists a large enough $\kappa_{0}$ such that $f(\kappa) < \frac{1}{p(\kappa)}$ for all $\kappa \geq \kappa_{0}$. We use the shorthand $f(\kappa) \geq \nonnegl(\kappa)$ to say that $f$ in non-negligible. We say that $f$ is super-logarithmic if $f(n)=\omega(\log n)$. For a boolean circuit $C$, \nom{$|C|$}{The number of gates in the circuit $C$} denotes the number of gates in it.

\begin{definition}[Collision resistant hashing scheme, adapted from~\cite{goldreich04foundations}]
\label{def:hashing_scheme}
Let $r(\kappa):\NN \to \NN$. A collection $\{h_{s}:\{0,1\}^{*}\to \{0,1\}^{r(|s|)}\}_{s \in \{0,1\}^{*}}$\nomenclature{$s$}{The index of a hash function}\nomenclature{$r$}{The range specifier of a hashing scheme. Also the length of a length-restricted TS scheme (see Definition~\ref{def:length_restricted_TS})} is called a collision resistant hashing scheme against quantum adversaries if there exists a PPT algorithm $\hindex$ such that the following holds:
\begin{enumerate}
\item There exists a polynomial time algorithm, that given $s$ and $x$, returns $h_{s}(x)$.
\item We say that the pair $x \neq x'$ forms a collision under the function $h$ if $h(x)=h(x')$. We require that for every QPT $\Adv$, given $\hindex(1^{\kappa})$ as an input, outputs a collision under $h_{\hindex(1^{\kappa})}$ with negligible probability:
\[ \Pr\left[\Adv(\hindex(1^{\kappa})) \text{ is a collision under } h_{\hindex(1^{\kappa})}\right] \leq \negl(\kappa)\]\item For some polynomial $p$, all sufficiently large $\kappa$'s, and every $s$ in the range of $\hindex(1^{\kappa})$, it holds that $\kappa \leq p(|s|)$. Furthermore, $\kappa$ can be computed in polynomial time from $s$.\footnote{ The main purpose of this technical requirement is to allow the adversary to run in time which is polynomial in $\kappa$.}
 
\end{enumerate}
\end{definition}
\begin{remark}
Our only use of hash-functions is in the context of the hash-and-sign paradigm, which extends a length-restricted scheme to an unrestricted one. Ref.~\cite{BSS21} shows that, as in the classical setting, one can replace the collision-resistant hash-function with a universal one-way hash functions. The analysis requires only modest modifications. The advantage of their approach is that all the cryptographic primitives, except the VBB obfuscation which we use (and they do not), follow generically from the existence of quantum secure one-way functions.  
\end{remark}

\begin{definition}[Digital Signature Scheme]\label{def:digital_signature}
	A digital signature (\nom{DS}{Digital Signature}) scheme consists of 3 PPT algorithms $\keygen,\ \sign$ and $\verify$ satisfying:
	\begin{enumerate}
		\item When a document is signed using the private key, the signature
		is accepted by the verification algorithm using the public key. I.e., for every $\alpha\in \{0,1\}^{*}$:
		\begin{equation}
		\Pr\left[\verify_{pk}\left(\alpha,\sign_{sk}\left(\alpha\right) \right)= 1 \right] = 1
		\label{eq:ds_correctness}
		\end{equation}
		\item The scheme is existentially unforgeable under chosen message
		attacks (EU-CMA): a quantum adversary with the capability of adaptively requesting
		documents to be signed by a classical signing oracle
		cannot generate a \emph{fresh} signed document. Formally,
		\begin{equation}
		\Pr\left[
		(\alpha,sig) \gets \Adv^{\sign_{sk}}(pk):\ \verify_{pk}(\alpha,sig) = 1 \text{ and } \alpha \notin Q_{\Adv}^{\sign_{sk}}\right] \leq \negl(\kappa),
        \label{eq:unforgeability_digital_signature} 
		\end{equation} 
		where $\Adv^{\sign_{sk}}$ is a QPT algorithm with access to the signing
		oracle, and $Q_{\Adv}^{\sign_{sk}}$ is the set of queries it made to the oracle.
	\end{enumerate}
	Whenever $\sign$ is a QPT algorithm (rather then PPT) we will say the scheme has \emph{quantum signing}.
\end{definition}

\begin{remark}
We will always assume (as we did in a Eq.~\eqref{eq:unforgeability_digital_signature}) that the public key $pk$ and the secret key $sk$ are sampled according to $\keygen(1^{\kappa})$ of the relevant protocol, and that $\ket{\stamp}$ is the output of $\tokengen_{sk}$ (to be defined later) and the probabilities always take that into account. 
\end{remark}
In the above definition, the probabilities range over
$(pk,sk) \gets \keygen(1^{\kappa})$ as well as the internal randomness
of all the algorithms. This will also be the case in all our future definitions.

We say a digital signature scheme is deterministic if $\verify$ is deterministic.
We also define a \emph{memory-dependent signature
	scheme (MDS)}. Such a scheme is defined similarly to the above,
except the chosen messages of the adversary must all be unique.\footnote{We define an MDS in a narrower sense than Goldreich's definition, which is sometimes called stateful signature scheme -- see~\cite[Definition 6.4.13]{goldreich04foundations}.}
It is clear that every DS is an MDS. The difference comes down to a technicality;
it is generally easier to prove a scheme is an MDS than a DS,
and it provides almost as much security (the signing protocol
of an MDS scheme can be made secure against all chosen
message attacks if it had memory, allowing it to respond to two signing
requests for the same document with the same signature). 



\section{Definitions}
\label{sec:definitions}

\begin{definition}[Public quantum money scheme
\protect\footnote{Also known as locally verifiable quantum money~\cite{mosca2010quantum}}
]
  A public quantum money scheme consists of three QPT algorithms:
  \begin{enumerate}
  \item $\keygen,$ which takes as input the security parameter $1^{\kappa}$ and probabilistically generates a key pair $pk,sk$.
   \item $\tokengen$, which takes as an input $sk$ and generates a quantum state $\$ $ called a (bank) token.
   \item $\verifytoken$  which takes as input $pk$ and an alleged bank token $\tau$, and outputs the post-measurement quantum state, and a classical output which is either accept or reject.  
  \end{enumerate}
$\verifytoken$ should accept a valid bank tokens $\$ $. Formally, for every $\kappa$:
\begin{equation}
    \Pr\left[ \$ \gets \tokengen_{sk}:  \verifytoken_{pk}(\$)=1 \right]= 1.
    \label{eq:qm_correctness}
\end{equation} 

Let $\verifytoken_{k,pk}$ takes as an input a collection of (possibly entangled) $k$ alleged bank tokens $\tau_{1},\ldots, \tau_{_{k}}$ and accepts if and only if for every $1 \leq i \leq k$, $\verifytoken_{pk}(\tau_{i})$ accepts. We say that the scheme is unforgeable if for every QPT $\Adv$:
\footnote{Aaronson and Christiano used a slightly different definition described next. Their definition is as follows. Let $\cnt$ take as input $pk$, as well as a collection of (possibly-entangled) alleged tokens $\tau_{1},\ldots,\tau_{\ell}$ and output the number of indices $i\in [r]$ such that $\verifytoken_{pk}(\tau_{i})$ accepts. We say that the scheme is secure if for every QPT $\Adv$ which maps the public key and $\ell = \poly(\kappa)$ valid tokens $\$_{1},\ldots,\$_{\ell}$ to $m$ (possibly entangled) alleged tokens $\tau_{1},\ldots,\tau_{m}$,
\begin{equation}
  \Pr\left[\cnt(\tau_{1},\ldots,\tau_{m}) > \ell \right] \leq \negl(\kappa).
\end{equation}
Note that the definition coincide for schemes which are secure against sabotage: The adversary could test whether verification passes using the public key, and only if it passes, the adversary will submit the resulting bill. If there is a discrepancy between the number of bills that were verified by the adversary, and the outcome of $\cnt$, then that same adversary could be used to violate the security against sabotage (see Eq.~\eqref{eq:sabotage_qm}).
}
\begin{equation}
  \Pr\left[\verifytoken_{\ell+1,pk}(\Adv(pk,\ket{\stamp_1}\tensor \ldots \tensor \ket{\stamp_\ell})) = 1 \right] \leq \negl(\kappa)
\label{eq:unforgeability_qm}
\end{equation}

The scheme is secure against sabotage if for every QPT $\Adv$ with access to $pk$ and polynomially many quantum money states, which generates a state $\sigma$:
\begin{equation}
  \Pr\left[
  (1,\tau)\gets\verifytoken_{pk}(\sigma) \text { and } \verifytoken_{pk}(\tau)=0  
  \right]\leq \negl(\kappa)
\label{eq:sabotage_qm}
\end{equation}
A scheme is secure if it is unforgeable and secure against sabotage.
\label{def:quatnum_money}
\end{definition}
\begin{remark}
To motivate the notion of sabotage in Eq.~\eqref{eq:sabotage_qm} (which had not appeared previously, to our knowledge), consider an adversary which aims to harm a specific merchant. Without the above requirement, one possible attack for the adversary (and there are other attacks with similar flavor) is to distort a quantum money state so that it will pass the verification test once, and fail on the second attempt of verification. How is that harmful? The adversary could use such a state to buy goods from the merchant. The merchant will run the verification, and assuming it accepts, the merchant will provide the goods to the adversary. When the merchant will try to spend this money, the verification would fail!
Eq.~\eqref{eq:sabotage_qm} guarantees that if a state passes the verification, it will continue to pass further verifications (except with a negligible probability).

Aaronson and Christiano used a projective scheme which guarantees that Eq.~\eqref{eq:sabotage_qm} holds (see~\cite{aaronson13quantum} for the definition of a projective scheme), but they did not consider it as part of the security requirement for quantum money.    
\label{rem:strong_security}
\end{remark}

\begin{definition}[Tokenized signature scheme]
\label{def:tokenized_signature_scheme}
A tokenized signature (\nom{TS}{Tokenized Signature scheme}) scheme consists of 4 algorithms, \keygen\ (PPT), \tokengen\ (QPT), \sign\ (QPT), and \verify\ (deterministic polynomial time) with the following syntax:
\begin{enumerate}
\item On input $1^{\kappa}$ where $\kappa$ is the security parameter, $\keygen$ outputs a classical public key $pk$ and a classical secret key $sk$.
\item $\tokengen_{sk}$ generates a signing token $\ket{\stamp}$. We emphasize that if $\tokengen_{sk}$ is called $\ell$ times it may (and in our construction, would) output different states $\ket{\stamp_{1}},\ldots,\ket{\stamp_{\ell}}$.
\item $\sign$ receives a quantum state (presumably, a signing token) and a document, and outputs a classical string, which is called a signature.
\item $\verify$ receives the public key, a document and an alleged signature and either accepts or rejects.
\end{enumerate}
The correctness property requires that for every document $\alpha \in \{0,1\}^{*}$, 
\begin{equation}
     \Pr\left[\ket{\stamp}\gets \tokengen_{sk}; sig \gets \sign(\alpha,\ket{\stamp}):\verify_{pk}(\alpha,sig)=1 \right]=1.
     \label{eq:ts_correctness}
\end{equation}
In an imperfect scheme, we replace that this event occurs (only) with overwhelming probability:
\begin{equation}
\Pr\left[\ket{\stamp}\gets \tokengen_{sk}; sig \gets \sign(\alpha,\ket{\stamp}):\verify_{pk}(\alpha,sig)=1 \right]\geq 1 - \negl(\kappa) .
\label{eq:imperfect_ts} 
\end{equation}
\end{definition}
All the results in this paper can be easily extended to imperfect schemes --- this is usually omitted to improve the presentation. Our main construction is imperfect.


\begin{definition}[Length restricted TS scheme] A TS scheme is $r$-restricted if the correctness property (see Eq.~\eqref{eq:ts_correctness}) holds only for $\alpha \in \{0,1\}^{r}$.
\label{def:length_restricted_TS}
\end{definition}

To define unforgeability, we introduce the algorithm \nom{$\verify_{k,pk}$}{Verification of $k$ documents, using the public key $pk$}. This algorithm takes as an input $k$ pairs $(\alpha_{1},s_{1}),\ldots,(\alpha_{k},s_{k})$ and accepts if and only if 
\begin{enumerate}
    \item \label{it:verify_k_1} All the documents are distinct, i.e. $\alpha_{i}\neq \alpha_{j}$ for every $1 \leq i \neq j \leq k$.
    \item \label{it:verify_k_2} All the pairs pass the verification test $\verify_{pk}(\alpha_{i},s_{i})$.
\end{enumerate}




\begin{definition}[Unforgeability and one-time unforgeability]
	\label{def:unforgeability}
A TS scheme is unforgeable if for every $\ell=\poly(\kappa)$ a QPT adversary cannot sign $\ell+1$ different documents by using the public key and $\ell$ signing tokens:
\begin{equation}
\Pr\left[\verify_{\ell+1,pk}( \Adv(pk,1^\kappa,\ket{\stamp_{1}} \tensor \ldots \tensor \ket{\stamp_{\ell}})) = 1 \right]\leq \negl(\kappa) 
\label{eq:unforgeability}
\end{equation}
One-time unforgeability requires the above only for $\ell=1$.
\end{definition}

\begin{remark}
Note that we assume that $\verify$ is deterministic. The main advantage of this syntax is that we do not need to consider sabotage attacks. If probabilistic or quantum verification is allowed, then one needs to add security against sabotage, along the line of Eq.~\eqref{eq:sabotage_qm}.
\end{remark}

\begin{remark}
In standard EUF-CMA security (see Definition~\ref{def:digital_signature}), the adversary is given access to a signing oracle, whereas here, the adversary is given signing tokens instead. What happens if we also augment a signing oracle (where the signing applies $\keygen$ and signs the requested message)? These definitions are equivalent as long as the the adversary is not allowed to sign the same document twice. If we do allow that, than the definitions are equivalent under  slightly stronger assumptions regarding the TS scheme called super-security and unpredictability --- see Remark~\ref{rem:ts_with_oracle} in  Appendix~\ref{app:super-secure}.
\end{remark}

We define $\revoke_{k,pk}$ in a similar way to $\verify_{k,pk}$: $\revoke_{k,pk}$  receives as an input $k$ registers, preforms $\revoke_{pk}$ on each one of them, and accepts iff all the $k$ revocations accept.

\begin{definition}[Revocability]
  A revocable tokenized signature scheme adds a fifth algorithm, $\revoke$ which satisfies:
\begin{enumerate}
\item $\Pr\left[\revoke_{pk}(\tokengen_{sk})=1 \right] = 1$.
\item  For every $\ell \leq \poly(\kappa)$, $t \leq \ell$, and every QPT $\Adv$ with $\ell$ signing tokens $\ket{\stamp_{1}}\tensor \ldots \tensor \ket{\stamp_{\ell}}$  and $pk$ which generates $\alpha_{1},sig_{1},\ldots, \alpha_{t},sig_{t}$ and a quantum state $\sigma$,
\[\Pr\left[\verify_{t,pk}(\alpha_{1},sig_{1},\ldots,\alpha_{t},sig_{t})=1 \ \wedge \  \revoke_{\ell-t+1,pk}(\sigma)=1\right] \leq \negl(\kappa).\]  
For one-time  schemes, we require the above only for $\ell=1$. 
\end{enumerate}
In imperfect schemes, we relax item 1 so that the r.h.s. is at least $1-\negl(\kappa)$. 
\label{def:revocable}
\end{definition}
We will show later in Theorem~\ref{thm:revocability} that every unforgeable TS scheme is revocable. 

Next, we define everlasting revocability. For a motivation, see the discussion in Section~\ref{sec:toekns_for_digital_signatures}.
\begin{definition}[One-time Everlasting Revocability] A revocable TS scheme has onetime everlasting revocability if a computationally unbounded quantum adversary $\Adv$ with access to 
$\ket{\stamp}$ (but \emph{without} access to $pk$), that generates 
a quantum state $\rho$, a document $\alpha$ and an alleged signature $sig$,
\begin{equation}
  \label{eq:everlasting_security}
  \Pr\left[ \revoke_{
  pk}(\rho)=1 \text{ and } \verify_{pk}(\alpha,sig)=1 \right] \leq \negl(k).
\end{equation}
\label{def:everlasting_security}
\end{definition}

A testable TS scheme has an additional algorithm $\verifytoken$, which, unlike $\revoke$, does not consume the signing token. If a state passes this test, it can be used to sign a document. 
\begin{definition}[Testability]
A testable TS scheme adds a fifth QPT algorithm to a TS scheme, $\verifytoken$, which satisfies:
\begin{enumerate}
\item \label{it:testability_correctness}$\Pr\left[\ket{\stamp} \gets \tokengen_{sk}: \verifytoken_{pk}(\ket{\stamp}) = 1 \right]=1$.
\item  For every QPT $\Adv$ with access to $pk$ and $\poly(\kappa)$ many signing tokens, which generates $\alpha$ and a state $\tau$:
\begin{equation}
\Pr\left[(a,\sigma) \gets \verifytoken_{pk}(\tau);b\gets \verify_{pk}(\alpha,\sign(\alpha,\sigma)): a = 1 \text{ and } b=0\right] \leq \negl(\kappa)
\label{eq:security_of_testability1}
\end{equation}
\begin{equation}
\Pr\left[(a,\sigma) \gets \verifytoken_{pk}(\tau); (b,\sigma')\gets \verifytoken_{pk}(\sigma):a=1 \text{ and } b=0\right] \leq \negl(\kappa)
\label{eq:security_of_testability2}
\end{equation}
The requirements in Eqs.~\eqref{eq:security_of_testability1} and \eqref{eq:security_of_testability2} are added for the same reasons as in Remark~\ref{rem:strong_security}. For one-time  schemes, the adversary is given a single signing token (rather than $\poly(\kappa)$ signing tokens).
\end{enumerate}
\label{def:testability}
\end{definition}

\section{Properties of Tokenized Signature Schemes}
\label{sec:properties-tokenized-signatures}
In this section, we prove several properties of tokenized signature schemes.

\begin{theorem}
  Every unforgeable tokenized signature scheme is also revocable (see Definition~\ref{def:revocable}) by the algorithm $\revoke$ given in Algorithm~\ref{alg:revoke}.
\label{thm:revocability}
\end{theorem}
The theorem does not hold for length-restricted schemes: as can be seen in Algorithm~\ref{alg:revoke}, the scheme must support signing messages of super-logarithmic length. 
\begin{algorithm}
    \caption{$\revoke$}
    \begin{algorithmic}[1] 
        \Procedure{$\revoke$}{$pk,\sigma$}
            \State $\alpha \getsr \{0,1\}^{f(\kappa)}$\Comment{$f$ can be any super-logarithmic function.}
            \State \textbf{return} $\verify_{pk}(\alpha, \sign(\alpha,\sigma))$
        \EndProcedure

    \end{algorithmic}
\label{alg:revoke}
\end{algorithm}

\begin{proof}
Assume towards a contradiction that $\revoke$ given in Algorithm~\ref{alg:revoke} does not satisfy Definition~\ref{def:revocable}, i.e. there exists  $\ell = \poly(\kappa)$ and $t$, and a QPT \Adv\ which generates $\alpha_{1},sig_{1},\ldots,\alpha_{t},sig_{t}$ and a state $\sigma$ such that
\begin{equation}
\Pr\left[ \verify_{t,pk}(\alpha_{1},sig_{1},\ldots,\alpha_{t},sig_{t})=1 \ \wedge \  \revoke_{\ell-t+1,pk}(\sigma)=1\right] \geq \nonnegl(\kappa).
\end{equation}

We define $\Adv'$ which uses $\Adv$ and simulates $\revoke$, which can be used to break the unforgeability of the $\ts$ scheme.
The adversary $\Adv'$ starts by running $\Adv$, and then uses the state $\sigma$ to produce signatures for additional $\ell$ random documents, see Algorithm~\ref{alg:adv_for_revoke}.

\begin{algorithm}
    \caption{$\Adv'$: the adversary used in the proof of Theorem~\ref{thm:revocability}}
    \begin{algorithmic}[1] 
        \Procedure{$\Adv'$}{$pk,\ket{\protect\stamp_{1}}\tensor \ldots \tensor \ket{\protect \stamp_{\ell}} $}
            \State $(\alpha_{1},sig_{1}, \ldots , \alpha_{t},sig_{t},\sigma_{t+1},\ldots,\sigma_{\ell+1}) \gets \Adv(pk,\ket{\stamp_{1}}\tensor \ldots \tensor \ket{ \stamp_{\ell}})$
            \ForAll{$i \in \{t+1, \ldots, \ell + 1\}$} 
                \State $\alpha_{i}\getsr \{0,1\}^{f(\kappa)}$ \Comment{$f$ can be any super-logarithmic function .}
                \State $sig_{i} \gets \sign(\alpha_{i},\sigma_{i})$
            \EndFor
            \State \textbf{return} $(\alpha,sig)$
        \EndProcedure

    \end{algorithmic}
\label{alg:adv_for_revoke}
\end{algorithm}

 By construction,

\begin{align}
 \Pr[&\verify_{\ell+1,pk}\left( \Adv'(pk,\ket{\stamp_{1}} \tensor \ldots \ket{\stamp_{\ell}})\right) = 1] \\
&= \Pr\left[ \verify_{t,pk}(\alpha_{1},sig_{1},\ldots,\alpha_{t},sig_{t})=1 \ \wedge \  \revoke_{\ell-t+1,pk}(\sigma)=1 \wedge \forall i\neq j,  \ \alpha_{i} \neq \alpha_{j}\  \right]   \\
&\geq  \Pr\left[ \verify_{t,pk}(\alpha_{1},sig_{1},\ldots,\alpha_{t},sig_{t})=1 \ \wedge \  \revoke_{\ell-t+1,pk}(\sigma)=1 \right] - \binom{\ell+1}{2} 2^{-f(\kappa)}  \\
&\geq \nonnegl(\kappa)
\label{eq:7}
\end{align}

which contradicts unforgeability (see Definition~\ref{def:unforgeability}).
\end{proof}

Next, we show that a variant of a tokenized signature scheme can be used as a digital signature scheme. 

Let $\ts$ be a tokenized signature scheme. We define a digital signature with quantum signing, denoted $\ds$, which uses $\ts$ in a black box manner. In $\ds$, $\keygen$ and $\verify$ are the same as in $\ts$, and $\ds.\sign_{sk}(\alpha)$ is $\ts.\sign(\alpha,\ts.\tokengen_{sk})$.


\begin{theorem}
  If $\ts$ is a tokenized signature scheme then the scheme $\ds$ as defined above is a memory-dependent digital signature with quantum signing (see Definition~\ref{def:digital_signature}).
\label{thm:ts_implies_mds}
\end{theorem}

The construction above only shows a memory-dependent digital signature scheme. 
In Theorem~\ref{thm:ts_implies_ds} in Appendix~\ref{app:super-secure} we show that memory-dependent property in the above construction can be omitted, given additional technical assumptions about the tokenized signature scheme. These additional assumptions hold for our main construction --- see Appendix~\ref{app:proving_super_security_and_unprdictability}.

\begin{proof}
It is clear that a valid signature will be accepted by this scheme, so we only
need to prove security. Suppose there was an adversary who used 
$\ell$ adaptive queries to the signing oracle and produced a signature for a fresh document, violating Eq.~\eqref{eq:unforgeability_digital_signature}.
Then this adversary produced valid signatures for $\ell+1$ different documents
using only $\ell$ calls to the oracle. Here we used the property that the documents given for the signing oracle are distinct, as required in an MDS (see Definition~\ref{def:digital_signature}). This means one can produce
$\ell+1$ valid signatures of distinct documents using only $\ell$ signing tokens, contradicting
the security assumption of the tokenized signature scheme.
\end{proof}

\begin{theorem}
  Every testable (see Definition~\ref{def:testability}) tokenized signature scheme is a public quantum money scheme (see Definition~\ref{def:quatnum_money}).
\label{thm:testable_ts_is_qm}
\end{theorem}


\begin{proof}
By Definition~\ref{def:testability}.\ref{it:testability_correctness} a testable tokenized signature scheme already has the correctness property of quantum money (see Eq.~\ref{eq:qm_correctness}), Eq.~\eqref{eq:security_of_testability1} guarantees security against sabotage (Eq.~\eqref{eq:sabotage_qm}). The only property which is left to show is unforgeability Eq.~\eqref{eq:unforgeability_qm}. Suppose towards a contradiction that unforgeability does not hold, i.e., there exists a QPT $\Adv$ which maps $pk,\ket{\stamp_{1}},\dots,\ket{\stamp_{\ell}}$, for $\ell=\poly(\kappa)$ to a state $\tau$ with $\ell+1$ registers such that 
\begin{equation}
  \Pr\left[ \verifytoken_{\ell+1,pk}(\tau) = 1 \right] \geq \nonnegl(\kappa).
\label{eq:1}
\end{equation}
\end{proof}

Let $\Adv'$ be the adversary as defined in Algorithm~\ref{alg:adv_for_tokenized_signatures_implies_qm}.
\begin{algorithm}
    \caption{$\Adv'$: the adversary used in the proof of Theorem~\ref{thm:testable_ts_is_qm}}
    \begin{algorithmic}[1] 
        \Procedure{$\Adv'$}{$pk,\ket{\protect\stamp_{1}}\tensor \ldots \tensor \ket{\protect \stamp_{\ell}} $}
            \State $\tau \gets \Adv(pk,\ket{\stamp_{1}} \tensor \ldots \tensor \ket{\stamp_{\ell}})$
            \If{$\verifytoken_{\ell+1,pk}(\tau) = (1,\sigma_{1},\ldots,\sigma_{\ell+1})$} 
                \ForAll{$i \in [\ell+1]$}
                    \State $\alpha_{i} \gets i$
                    \State $sig_{i} \gets \sign(\alpha_{i},\sigma_{i})$
                \EndFor
            
            \EndIf
            \State \textbf{return} $(\alpha,sig)$
        \EndProcedure

    \end{algorithmic}
\label{alg:adv_for_tokenized_signatures_implies_qm}
\end{algorithm}

\begin{align}
  \Pr&\left[ \verify_{\ell+1,pk}(\Adv'(pk,\ket{\stamp_{1}}\tensor\ldots \tensor \ket{\stamp_{\ell}})) = 1\right] \\
 &=\Pr\left[\verifytoken_{\ell+1,pk}(\tau) = (1,\sigma) \wedge \verify_{\ell+1,pk}(\alpha,sig) = 1 \right] \\
&=\Pr\left[\verifytoken_{\ell+1,pk}(\tau) = (1,\sigma) \bigwedge_{i \in [\ell + 1]} \verify_{pk}(\alpha_{i},\sign(\alpha_{i},\sigma_{i})) = 1\right] \\
&\geq \nonnegl(\kappa)
\label{eq:10}
\end{align}
where in the last step we used Eq.~\eqref{eq:1}  and Eq.~\eqref{eq:security_of_testability1}. Eq.~\eqref{eq:10} contradicts unforgeability (Definition~\ref{def:unforgeability}). 
\section{Lifting a 1-bit One-time Scheme}
\label{sec:standard-construction}
In this section we show how a 1-bit length restricted (see Definition~\ref{def:length_restricted_TS}) one-time  (see Definition~\ref{def:unforgeability})  tokenized signature scheme, denoted $\ot1$, can be lifted to a full-fledged tokenized signature scheme (i.e used to sign an arbitrary number of documents, of arbitrary length). 

This is done using three reductions, which construct three additional schemes. The second scheme, $\otr$ (see Algorithm~\ref{alg:otr}), is one-time $r$-restricted: it is achieved by using the $\ot1$ scheme $r$ times. The third scheme, $\ot$ (see Algorithm~\ref{alg:ot}), removes the length restriction by using the hash-and-sign paradigm: instead of signing the document $\alpha$ (which might be arbitrarily long), we sign the document $h_{s}(\alpha)$ where $h_{s}$ is a cryptographic hash function. The proof relies on the assumption that $h$ is post-quantum collision resistant. The proof is the same as the classical security proof of the hash-and-sign paradigm (see, for example,~\cite{goldreich04foundations}), and can be safely skipped by those familiar with it. 
The fourth scheme, $\ts$ (see Algorithm~\ref{alg:ts}), uses a (quantum resistant) digital signature scheme (denoted $\ds$). The public and secret keys are those generated by the digital signature. To mint, we first run the $\ot$ key-generation to produce $(sk',pk')$, use $sk'$ to mint a quantum money state, sign $pk'$ using $sk$ and append $pk'$ and its signature to the quantum money. Verification is done by first checking that $pk'$ is signed correctly, and using $pk'$ to verify the quantum money. The fourth construction is very similar to the mini-scheme to full-scheme construction in~\cite{aaronson13quantum}.

In rest of this section we show that all the reductions maintain unforgeability (see Definition~\ref{def:unforgeability}), testability (Definition~\ref{def:testability}), and that the first two reductions (but not the last) maintain  everlasting revocability (see Definition~\ref{def:everlasting_security}). We will always assume that $r \leq \poly(\kappa)$. 

\begin{algorithm}
    \caption{$\otr$: One-time $r$-restricted scheme}
   
    \begin{algorithmic}[1] 
        \Procedure{$\keygen$}{$1^{\kappa},1^{r}$}
            \ForAll{$i \in [r]$}
                \State $(pk_{i},sk_{i})\gets \ot1.\keygen(1^{n})$
            \EndFor
            \State \textbf{return} $(pk=(pk_1,\ldots,pk_r),sk=(sk_1,\ldots,sk_r))$
        \EndProcedure

        \Procedure{$\tokengen$}{$sk$}
            \ForAll{$i \in [r]$}
                \State $\ket{\stamp_{i}}\gets \ot1.\tokengen(sk_{i})$
            \EndFor
            \State \textbf{return} $\ket{\stamp}=\ket{\stamp_{1}}\tensor \ldots \tensor \ket{\stamp_{r}}$

        \EndProcedure

        \Procedure{$\sign$}{$\alpha \in \{0,1\}^{r},\ket{\protect \stamp}$}
            \ForAll{$i \in [r]$}
                \State $sig_{i}\gets \ot1.\sign(\alpha_{i},\ket{\stamp_{i}})$
            \EndFor
            \State \textbf{return} $sig=(sig_1,\ldots,sig_r)$
        \EndProcedure

        \Procedure{$\verify_{pk}$}{$\alpha \in \{0,1\}^{r}, sig$}
                \If{\textbf{for all } $i \in [r],\, \ot1.\verify(\alpha_{i},sig_{i})=1$} 
                    \State \textbf{return} 1
                \Else
                    \State \textbf{return} 0
                \EndIf
                    
        \EndProcedure
        \Procedure{$\verifytoken_{pk}$}{$\tau$}
            \If{\textbf{for all } $i \in [r],\ \ot1.\verifytoken_{pk_{i}}(\tau_{i})=(1,\sigma_{i})$} 
                \State \textbf{return} $(1, \sigma)$ 
            \Else
                \State  \textbf{return} $(0,\sigma)$ 
            \EndIf

        \EndProcedure
 
    \end{algorithmic}
\label{alg:otr}
\end{algorithm}

\begin{algorithm}
    \caption{$\ot$: one-time scheme}
    \begin{algorithmic}[1] 
        \Procedure{$\keygen$}{$1^{\kappa}$}
            \State $s \gets \hindex(1^{\kappa})$ \Comment{See Definition~\ref{def:hashing_scheme} for $\hindex$.}
            \State  $(pk',sk') \gets \otr.\keygen(1^\kappa,1^{r(\kappa)})$ \Comment{See Definition~\ref{def:hashing_scheme} for $r(\kappa)$.}
            \State \textbf{return} $(pk=(pk',s),sk=(sk',s))$
        \EndProcedure

        \Procedure{$\tokengen$}{$sk$}
            \State $\ket{\stamp'} \gets \otr.\tokengen(sk')$
            \State \textbf{return} $\ket{\stamp}=(s,\ket{\stamp'})$  \Comment{We add $s$ to the token since $\sign$ does not have access to $pk$.}
        \EndProcedure

        \Procedure{$\sign$}{$\alpha \in \{0,1\}^{*},(s,\ket{\protect \stamp}$)}
            \State \textbf{return} $\otr.\sign(h_{s}(\alpha),\ket{\stamp})$
        \EndProcedure
        
        \Procedure{$\verify_{pk}$}{$\alpha \in \{0,1\}^{*}, sig$}
            \State \textbf{return} $\otr.\verify_{pk'}(h_{s}(\alpha),sig)$     
        \EndProcedure

        \Procedure{$\verifytoken_{pk}$}{$s', \tau$}
            \If{$s=s'$ and $\otr.\verifytoken_{pk}(\tau)=(1,\sigma)$} 
                \State \textbf{return} $(1, (s, \sigma))$ 
            \Else
                \State  \textbf{return} $(0,(s,\sigma))$ 
            \EndIf

        \EndProcedure
    \end{algorithmic}
\label{alg:ot}
\end{algorithm}

\begin{algorithm}
    \caption{$\ts$: Tokenized Signature scheme}
   
    \begin{algorithmic}[1] 
        \Procedure{$\keygen$}{$1^{\kappa}$}
            \State  $(pk,sk) \gets \ds.\keygen(\kappa)$
            \State \textbf{return} $(pk,sk)$
        \EndProcedure

        \Procedure{$\tokengen$}{$sk$}
            \State $(\ot.pk,\ot.sk) \gets \ot.\keygen(1^{\kappa},r)$
            \State $\ket{\stamp'} \gets \ot.\tokengen(\ot.sk)$
            \State $\ket{\stamp} \gets (\ot.pk,\ds.\sign_{sk}(\ot.pk),\ket{\stamp'})$ \label{line:tokengen_output}
            \State \textbf{return} $\ket{\stamp}$
        \EndProcedure

        \Procedure{$\sign$}{$\alpha \in \{0,1\}^{*},\ket{\protect \stamp}$}
            \State Assume $\ket{\stamp}$ has the form $(\ot.pk,\ds.\sign_{sk}(\ot.pk),\ket{\stamp'})$
            \State $sig' \gets \ot.\sign(\alpha,\ket{\stamp'})$
            \State \textbf{return} $sig= (\ot.pk,\ds.\sign_{sk}(\ot.pk) ,sig' )$ \label{line:ts_output_sign}
        \EndProcedure
        
        \Procedure{$\verify_{pk}$}{$\alpha \in \{0,1\}^{*}, sig$}
            \State Assume $sig$ has the form $(\ot.pk,\ds.\sign_{sk}(\ot.pk) ,sig' )$
            \State \textbf{return} $ \ds.\verify_{pk}(\ot.pk,\ds.\sign_{sk}(\ot.pk)) \wedge \ot.\verify_{\ot.pk}(\alpha,sig')$  \label{line:ot_verify_return}   
        \EndProcedure
        \Procedure{$\verifytoken_{pk}$}{$\tau$}
            \State Assume $\tau$ has the form $(\ot.pk,\ds.\sign_{sk}(\ot.pk),\tau')$
             \If{$\ds.\verify_{pk}(\ot.pk,\ds.\sign_{sk}(\ot.pk)) = 1$ and $\ot.\verifytoken_{\ot.pk}(\tau')=(1,\sigma) $}
                \State \textbf{return} $(1, (\ot.pk, \ds.\sign_{sk}(\ot.pk),  \sigma))$ 
            \Else
                \State  \textbf{return} $(0,(\ot.pk, \ds.\sign_{sk}(\ot.pk),  \sigma))$ 
            \EndIf
        \EndProcedure
    \end{algorithmic}
\label{alg:ts}
\end{algorithm}

\begin{theorem}

  \begin{enumerate}
  \item \label{it:standard_construction_1} If $\ot1$ is a $1$-restricted one-time unforgeable TS scheme, then $\otr$ is $r$-restricted one-time unforgeable TS scheme.
  \item  \label{it:standard_construction_2} If $\otr$ is $r$-restricted one-time unforgeable TS scheme and $\{h_{s}\}_s$ is a collision resistant hashing scheme, then $\ot$ is one-time unforgeable TS scheme.
  \item \label{it:standard_construction_3}  If $\ot$ is a one-time unforgeable TS scheme and $\ds$ is a EU-CMA digital signature scheme, then $\ts$ is a (full blown) unforgeable TS scheme. 
  \end{enumerate}
 \label{thm:standard_construction_unforgability}
\end{theorem}

\begin{proof} 
The correctness property in all the steps follows trivially by construction.

We start by proving item~\ref{it:standard_construction_1}. To show one-time unforgeability, we show how to map a successful attack given by $\Adv$ on $\otr$ to a successful attack $\Adv'$ on $\ot1$, hence a contradiction. The algorithm for $\Adv'$ is given in Algorithm~\ref{alg:adv_ot1}.
\begin{algorithm}
    \caption{$\Adv'$: the adversary used in the proof of Theorem~\ref{thm:standard_construction_unforgability}}
    \begin{algorithmic}[1] 
        \Procedure{$\Adv'$}{$pk,\ket{\protect \stamp}$}
            \State $j \getsr [r]$.
            \State $(pk_{j}',\ket{\stamp'_{j}}) \gets (pk,\ket{\stamp})$
            \ForAll {$i \in [r] \setminus \{j\}$}
                \State $(pk'_{i},sk'_{i}) \gets \ot1.\keygen(\kappa)$.
                \State $\ket{\stamp'_{i}} \gets \ot1.\tokengen(sk'_{i})$
            \EndFor
            \State $(\alpha_0,\alpha_{1},s_{0},s_{1}) \gets \Adv(\mathbf{pk'} ,\ket{\stamp'_{1}}\tensor \ldots \tensor \ket{\stamp'_{r}})$
            \If{$\alpha_{0}[j] \neq \alpha_{1}[j] $}
                \State \textbf{return } $(\alpha_0[j],\alpha_1[j],s_{0}[j],s_{1}[j])$ 
            \EndIf
        \EndProcedure
    \end{algorithmic}
\label{alg:adv_ot1}
\end{algorithm}
In this algorithm, $\Adv'$ prepares $r-1$ public keys and secret keys to simulate the additional public keys that are part of the $\otr$ scheme. Suppose the success probability of $\Adv$ is greater than a non-negligible function denoted $f(\kappa)$. The success probability  of $\Adv'$ is at least $\frac{f(\kappa)}{r}\geq \nonnegl(\kappa)$, contradicting the unforgeability assumption for $\ot1$:

\begin{align}
  &\Pr\left[ \ot1.\verify_{2,pk}\left(\Adv'(pk,\ket{\stamp})\right)= 1 \right] \\
\geq &\Pr\left[ \otr.\verify_{2,pk'}\left(\Adv(pk',\ket{\stamp_{1}}\tensor \ldots \tensor \ket{\stamp_{r}}) \right) = 1 \text{ and } \alpha_{0}[j] \neq \alpha_{1}[j]\right] \\
 \geq & \frac{1}{r} \Pr\left[ \otr.\verify_{2,pk'}\left(\Adv(pk',\ket{\stamp_{1}}\tensor \ldots \tensor \ket{\stamp_{r}})\right)=1\right] = \frac{f(\kappa)}{r}=\nonnegl(\kappa)
\end{align}

Here, the second inequality is justified by the fact that in order for $\otr.\verify_{2,pk'}$ to accept, $\alpha_{0}$ must differ from $\alpha_{1}$. Therefore, the probability that they differ on the random index $j$ may be smaller by at most a factor of $\frac{1}{r}$.

We now show item~\ref{it:standard_construction_2}.
The proof is essentially the same proof of the hash-and-sign paradigm for one-time length restricted digital signature scheme, see,  for example, ~\cite[Proposition 6.4.7]{goldreich04foundations}.  
The proof is straightforward and provided here for the sake of completeness: suppose the adversary can find two documents $\alpha_{1}$ and $\alpha_{2}$ and signatures $sig_{1}$ and $sig_{2}$ which both pass the verification for $\ot$ with non-negligible probability. There are two cases: if $h_{s}(\alpha_{1})=h_{s}(\alpha_{2})$, the adversary found a collision of the (collision-resistant) hashing scheme. By a standard argument, this can only occur with negligible probability by the security of the hashing scheme, so this case can be ignored. If $h_{s}(\alpha_{1})\neq h_{s}(\alpha_{2})$, then $(\alpha'_{1}\equiv h_{s}(\alpha_{1}),sig_{1})$ and $(\alpha'_{2}\equiv h_{s}(\alpha_{2}),sig_{2})$ are valid document-signature pairs, and violate the (assumed) unforgeability property of the scheme $\otr$ -- contradiction.

Next, we prove item~\ref{it:standard_construction_3}.
The proof is similar to the proof of the \emph{standard construction} of Aaronson and Christiano, which elevates a quantum money mini-scheme to a (full blown) public-key quantum money scheme. A sketch proof is provided here for completeness.

Suppose towards a contradiction that there exists a QPT $\Adv$ which uses $\ell$ signing tokens, $\ket{\stamp_{1}}\otimes \ldots \otimes \ket{\stamp_{\ell}}$ generated by $\ts.\tokengen$ to prepare $(\alpha_{1},sig_{1},\ldots,\alpha_{\ell+1},sig_{\ell+1})$ such that $\ts.\verify_{\ell+1,pk}(\alpha_{1},sig_{1},\ldots,\alpha_{\ell+1},sig_{\ell+1})$ accepts with non-negligible probability. Assume that each $sig_{i}$ has the form $p_{i}$, $\ds.\sign_{sk}(p_{i})$, $sig'_{i}$. Recall that each $\ket{\stamp_{i}}$ has the form $\ot.pk_{i}, \ds.\sign_{sk}(pk_{i}),\ket{\stamp'_{i}}$. Let $PK=\{\ot.pk_{i}\}_{i \in \ell}$. We claim that for each $i\in [\ell +1]$, $p_{i} \in PK$. This is justified since the first step in $\ot.\verify$ is to check for the validity of the digital signature scheme. The only documents for which the adversary has access to valid signatures are $PK$, and, therefore, creating any other signature would break the unforgeability property of $\ds$. 

By the pigeonhole principle, there exists $k \in [\ell]$ and $i \neq j \in [\ell + 1]$ such that $p_{i} = p_{j} = \ot.pk_{k}$. Furthermore,   $\ot.\verify_{2,\ot.pk_{k}} (\alpha_{i},sig'_{i} ,\alpha_{j},sig'_{j})$ accepts with non-negligible probability as part of the verification test (see line~\ref{line:ot_verify_return} in Algorithm~\ref{alg:ts}). Only a single signing token associated with $\ot.pk_{k}$ was given to the adversary (as part of $\ket{\stamp_{k}}$), and therefore, by a standard unforgeability argument we reach a contradiction to the one-time-unforgeability property of $\ot$.
\end{proof}

\begin{theorem}
  \begin{enumerate}
  \item If $\ot1$ is testable, then $\otr$ is testable.  \label{it:standard_construction_testability_1} 
  \item If $\otr$ is testable then $\ot$ is testable.\label{it:standard_construction_testability_2}
  \item If $\ot$ is testable and $\ds$ is a deterministic digital signature scheme then $\ts$ is testable. \label{it:standard_construction_testability_3} 

  \end{enumerate}
\label{thm:standard_construction_testable}
\end{theorem}
\begin{proof}
The correctness requirement in Definition~\ref{def:testability}.\ref{it:testability_correctness} follows trivially from the properties of the construction for (all) cases 1-3.

Proof of item~\ref{it:standard_construction_testability_1}.  We show that the requirement in Eq.~\eqref{eq:security_of_testability1} holds. 
For every QPT $\Adv$ with access to $pk$ and $\poly(\kappa)$ many signing tokens, which generates $\alpha$ and a state $\tau$:

\begin{align}
\Pr&\left[(a,\sigma) \gets \otr.\verifytoken_{pk}(\tau); b \gets \otr.\verify_{pk}(\alpha,\otr.\sign(\alpha,\sigma));a=1 \text{ and } b=0\right]\\
&=\Pr\left[\bigwedge_{i=1}^{r} (1,\sigma_{i}) \gets \ot1.\verifytoken_{pk_{i}}(\tau_{i}) \text{ and }\bigvee_{i=1}^{r}\ot1.\verify_{pk_{i}}(\alpha_{i},\ot1.\sign(\alpha_{i},\sigma_{i}))=0 \right]\\
&\leq \Pr\left[\bigcup_{i=1}^{r} (1,\sigma_{i}) \gets \ot1.\verifytoken_{pk_{i}}(\tau_{i}) \text{ and }\ot1.\verify_{pk_{i}}(\alpha_{i},\ot1.\sign(\alpha_{i},\sigma_{i}))=0 \right]\\
&\leq \sum_{i=1}^{r} \Pr\left[(1,\sigma_{i}) \gets \ot1.\verifytoken_{pk_{i}}(\tau_{i}) \text{ and }\ot1.\verify_{pk_{i}}(\alpha_{i},\ot1.\sign(\alpha_{i},\sigma_{i}))=0 \right]\\
&\leq r \cdot \negl(\kappa) = \negl(\kappa),
\end{align}
where we use the union bound and the assumption that $\ot1$ is testable in the last two inequalities, respectively. 

The requirement in Eq.~\eqref{eq:security_of_testability2} can be shown by following the same steps of the proof above (while replacing $\verify$ with $\verifytoken$ appropriately), and is therefore omitted.

Proof of item~\ref{it:standard_construction_testability_2}: immediate.

Proof of item~\ref{it:standard_construction_testability_3}: We show that the requirement in Eq.~\eqref{eq:security_of_testability1} holds. For every QPT $\Adv$ with access to $pk$ and $\poly(\kappa)$ many signing tokens, which generates a document $\alpha$ and a triple $(\ot.pk,sig,\tau)$,
\begin{align}
\Pr&\left[\begin{matrix}
(a,(\ot.pk,sig,\sigma)) \gets \ts.\verifytoken_{pk}(\ot.pk,sig,\tau)\\
b \gets \ts.\verify_{pk}(\alpha,\sign(\alpha,(\ot.pk,sig,\sigma)))
\end{matrix} :a=1 \text{ and } b=0\right]\\
&=\Pr\left[\begin{matrix}
a_0\gets \ds.\verify_{pk}(\ot.pk,sig) \\
(a_1,\sigma)\gets \ot.\verifytoken_{\ot.pk}(\tau)\\
b_0 \gets \ds.\verify_{pk}(\ot.pk,sig) \\
b_1 \gets \ot.\verify_{\ot.pk}(\alpha,\sign(\alpha,\sigma))
\end{matrix}
: (a_0=1 \wedge a_1=1) \wedge (b_0=0 \vee b_1=0) \right] \\
&=\Pr\left[\begin{matrix}
a_0\gets \ds.\verify_{pk}(\ot.pk,sig) \\
(a_1,\sigma)\gets \ot.\verifytoken_{\ot.pk}(\tau)\\
b_1 \gets \ot.\verify_{\ot.pk}(\alpha,\sign(\alpha,\sigma))
\end{matrix}
: a_0=1 \wedge a_1=1 \wedge  b_1=0 \right] \\
&\leq \Pr\left[\begin{matrix}
(a_1,\sigma)\gets \ot.\verifytoken_{\ot.pk}(\tau)\\
b_1 \gets \ot.\verify_{\ot.pk}(\alpha,\sign(\alpha,\sigma))
\end{matrix}
: a_1=1 \wedge  b_1=0 \right] \\
&\leq \negl(\kappa).
  \end{align}
In the second equality, we used the property that the verification of digital signatures is deterministic (see Definition~\ref{def:digital_signature}) and therefore $a_0=b_0$; in the last inequality, we used the security of the underlying scheme $\ot$: if the inequality does not hold, we can easily construct an adversary which violates ~\eqref{eq:security_of_testability2}. The requirement in Eq.~\eqref{eq:security_of_testability2} is shown by following the same reasoning  (while replacing $\verify$ with $\verifytoken$ appropriately), and is therefore omitted.
\end{proof}

\begin{theorem}
  \begin{enumerate}
  \item If $\ot1$ has one-time everlasting revocability (see Definition~\ref{def:revocable}), then so does $\otr$. \label{it:standard_construction_everlasting_1} 
  \item If $\otr$ has one-time everlasting revocability (see Definition~\ref{def:revocable}), then so does $\ot$.   \label{it:standard_construction_everlasting_2} 
  \end{enumerate}

\label{thm:standard_construction_everlasting}
\end{theorem}

\begin{proof}
Item~\ref{it:standard_construction_everlasting_1} follows trivially, since $\otr$ is simply a repetition of $\ot1$. 

Proof of item~\ref{it:standard_construction_everlasting_2}. We reduce an attack by $\Adv$ on $\ot$ to an attack by $\Adv'$ on $\otr$. $\Adv'$ repeats the same procedure as $\Adv$, except that the document which it outputs is $h_{s}(\alpha)$ instead of $\alpha$ ($\Adv'$ has access to $s$ as it is part of the signing token).  
\begin{align}
  \Pr&\left[\begin{matrix}(\rho,\alpha,sig)\gets  \Adv(\ket{\stamp})\\
  a \gets \ot.\verify_{pk}(\alpha,sig)\\
  b \gets \ot.\revoke_{pk}(\rho)
\end{matrix}
: a=1 \text{ and } b=1\right]\\
&= \Pr\left[\begin{matrix}(\rho,\alpha,sig)\gets  \Adv(\ket{\stamp})\\
  a \gets \otr.\verify_{pk'}(h(\alpha),sig)\\
  b \gets \otr.\revoke_{pk'}(\rho)
\end{matrix}
: a=1 \text{ and } b=1\right]\\
&= \Pr\left[\begin{matrix}(\rho,h(\alpha),sig)\gets  \Adv'(\ket{\stamp})\\
  a \gets \otr.\verify_{pk'}(h(\alpha),sig)\\
  b \gets \otr.\revoke_{pk'}(\rho)
\end{matrix}
: a=1 \text{ and } b=1\right]\\
&\leq \negl(\kappa), \qedhere
\end{align}
where the last inequality follows directly from the one-time everlasting revocability of $\otr$.
\end{proof}

\section{Security in the Oracle Model}
\label{sec:oracle-results}
In this section, we prove the security of the scheme relative to an oracle,
proving the following theorem. We will use the notation \nom{$\Lambda(A)$}{The set $A \setminus \{0\} \times A^\perp \setminus \{0\}$} to denote the target set $A \setminus \{0\} \times A^\perp \setminus \{0\}$.

\begin{restatable}{theorem}{LowerBoundAverageCase}	\label{thm:lower-bound-average-case}
	Let $A$ be a uniformly random subspace from $S(n)$,
	and let $\epsilon>0$ be such that $1/\epsilon=o\left(2^{n/2}\right)$.
	Given one copy of $\ket{A}$ and a quantum membership oracle for $A$ and $A^\perp$,
	a counterfeiter needs $\Omega(\sqrt{\epsilon} 2^{n/4})$ queries
	to output a pair $(a,b)\in \Lambda(A)$
	with probability at least $\epsilon$.
\end{restatable}

Our approach is similar to the inner-product adversary method of Aaronson
and Christiano. We start by reviewing the oracle model.

\subsection{The Oracle Model}

In the oracle model, we assume that we start with a single copy of the state $\ket{A}$
and have quantum query access to an oracle for membership in the spaces $A$
and $A^\perp$.

To provide quantum query access, we allow the algorithm to apply 
the reversible (unitary) version of the unified membership function $\IAH$ (see Eq.~\eqref{eq:IAH}). Formally,  
$U_{A^*}$ acts on a Hilbert space whose
basis states are triples $(a,x,p)$ with $a\in\mathbb{F}_2^n$ and $x,p\in\{0,1\}$;
we then define \nom{$U_{A^*}$}{A quantum unitary which can be used to query membership in $A$ and $A^\perp$} by
its action on these basis states, which we define to be 
\begin{equation}
    \ket{a,p,x} \to \ket{a,p,\IAH(a,p) \oplus x}.
\end{equation}
The interpretation of this is that the bit $p$
controls whether the access is to the $A$ oracle or the
$A^\perp$ oracle, $a$ dictates the vector whose membership is to be checked, and the bit $x$ stores the output
(for reversibility reasons, the output is XORed with $x$).
We will further extend this unitary $U_{A^*}$ to act
on a larger space with more registers; $U_{A^*}$
will simply act as identity on the other registers.

We wish to use a quantum algorithm to produce
two elements $(a,b)\in \Lambda(A)$.
The cost of the algorithm will be the number
of calls it makes to the oracle $U_{A^*}$.

We will think of a quantum algorithm as keeping a superposition
over four registers: the query register (specifying the next oracle call), a work register, and two output registers (which are measured at the end).
The query register has the form $\ket{a,p,x}$ where $a\in\mathbb{F}_2^n$
and $p,x\in\{0,1\}$; the unitary $U_{A^*}$ will be applied to this register, and it will act as identity 
on the other registers.
The work register has an arbitrary (finite) dimension and is initialized to a fixed
state $\ket{0}_W$. The two output registers get measured at the end of the
algorithm, and each stores an element of $\mathbb{F}_2^n$.
Since we want the algorithm to have access to one copy of $\ket{A}$,
we will initialize one of the output registers to $\ket{A}$ and the other to
$\ket{0}$. The initial state will therefore be
\[\ket{\psi^A_{init}}:=\ket{0,0,0}\ket{0}_W\ket{A}\ket{0}.\]

A quantum algorithm is fully described by a sequence of unitaries $U_0,U_1,\ldots,U_T$ acting on these registers.
Applying such an algorithm is achieved
by alternately applying the $U_i$ unitaries
and the query unitary $U_{A^*}$, starting from the initial state $\ket{\psi^A_{init}}$.
The state at the end of the algorithm is therefore
\[\ket{\psi^A_{f}}:=U_TU_{A^*}U_{T-1}U_{A^*}\ldots U_{A^*}U_1U_{A^*}U_0\ket{\psi_{init}^A}.\]
The two output registers are then measured, and the result constitutes
the output of the algorithm.

\subsection{Lower Bound for Worst-Case Bounded-Error Algorithms}


We will show a lower bound on the number of queries
required by an algorithm that succeeds in producing a pair
$(a,b)\in\Lambda(A)$ in the oracle model, where success
is with probability at least $0.99$ and where success
is achieved against worst-case $A\in S(n)$. To show this,
we will use the following theorem.

\begin{theorem}[Adapted from \cite{aaronson13quantum}]
	\label{thm:inner-product-adversary-method}
	Let $\mathcal{R}$ be a symmetric, anti-reflexive relation on $S(n)$ such that
    each space in $S(n)$ related to at least one other space.
    For any $A\in S(n)$ and any algorithm $(U_0,U_1,\dots,U_T)$,
    define $\ket{\psi_{init}^A}$ and $\ket{\psi_{f}^A}$ as above, so that
    $\ket{\psi^A_{f}}=U_TU_{A^*}\ldots U_{A^*}U_0\ket{\psi_{init}^A}$.
	Suppose that for some positive scalars $c$ and $d$,
	we have $|\braket{\psi_{init}^{A}}{\psi_{init}^{B}} |\geq c$
	for all $(A,B)\in \mathcal{R}$ and
	$\E_{(A,B) \in \mathcal{R}}[|\braket{\psi_{f}^{A}}{\psi_{f}^{B}}|] \leq d$.
	Then any algorithm achieving this must make a number of oracle queries $T$ which is at least
	$\Omega\left((c-d)2^{n/2}\right)$.
\end{theorem}

\begin{remark}
	Aaronson and Christiano claimed a slightly weaker result. Their assumption is
	$\forall {A,B} \in \mathcal{R}$, $|\braket{\psi_{f}^{A}}{\psi_{f}^{B}}| \leq d$, while ours is $\E_{(A,B) \in \mathcal{R} }[|\braket{\psi_{f}^{A}}{\psi_{f}^{B}}|] \leq d$. However, their original proof is valid for this stronger version.
\end{remark}

We fix $\mathcal{R} \subset S(n) \times S(n)$ \nomenclature{$\mathcal{R}$}{The relation that contains all $(A,B)$ such that $\dim(A \cap B)=\frac{n}{2} - 1$} to be the relation that contains all $(A,B)$ such that $\dim(A \cap B)=\frac{n}{2} - 1$. We use \nom{$\mathcal{R}_A$}{The set of $B\in S(n)$ such that $(A,B)\in\mathcal{R}$}
to denote the set of $B\in S(n)$ such that $(A,B)\in\mathcal{R}$.

Note that we have $|\braket{\psi_{init}^{A}}{\psi_{init}^{B}} |=|\braket{A}{B}|\geq 1/2$
for all $(A,B)\in\mathcal{R}$, because our choice of $\mathcal{R}$ means that
$A$ and $B$ agree on $n/2-1$ out of $n/2$ dimensions.
To apply Theorem~\ref{thm:inner-product-adversary-method}, we will upper bound
$\E_{(A,B) \in \mathcal{R}}[|\braket{\psi_{f}^{A}}{\psi_{f}^{B}}|]$ by $0.45$
for any algorithm which succeeds at outputting from $\Lambda(A)$ with high enough
probability.
this will give us a lower bound of $\Omega(2^{n/2})$ on the number of queries.


\begin{lemma} \label{lem:upper-bound-on-final-states}
	For $A\in S(n)$, let $\ket{\psi^{A}_{f}}$ be the final state of the algorithm when
	the oracle and initial state encode the space $A$.
	If the algorithm outputs a pair $(a,b)\in\Lambda(A)$ with probability
	at least $0.99$ for all $A\in S(n)$, then
	\[ \E_{(A,B) \in R}[|\braket{\psi^{A}_f}{\psi^{B}_f}|] \leq 
	0.2 + \max_{A \in S(n), (a,b)\in \Lambda(A)}
	\Pr_{B \sim \mathcal{R}_A}[(a,b) \in \Lambda(B)].\]
\end{lemma}
\begin{proof}
	We will decompose the final state $\ket{\psi_{f}^A}$ as
	\[\ket{\psi_{f}^A}=\sum_{a,b\in \mathbb{F}_2^n}\beta_{ab}^A\ket{\phi^A_{ab}}\ket{a}\ket{b}
	=\sum_{(a,b)\in \Lambda(A)}\beta_{ab}^A\ket{\phi^A_{ab}}\ket{a}\ket{b}
	+\sum_{(a,b)\notin\Lambda(A)}\beta_{ab}^A\ket{\phi^A_{ab}}\ket{a}\ket{b}.\]
	Note that $\sum_{a,b\in\mathbb{F}_2^n}|\beta_{ab}^A|^2=1$ and that
	$\sum_{(a,b)\notin \Lambda(A)}|\beta_{ab}^A|^2\leq 0.01$.
    Also, the two sums above form orthogonal vectors;
    denoting them $\ket{v_A}$ and $\ket{u_A}$, and similarly for $B$, we have
    $\ket{\psi_{f}^A}=\ket{v_A}+\ket{u_A}$ and $\|\ket{v_A}\|\leq 1$, $\|\ket{u_A}\|\leq 0.1$.
	It follows that 
    \begin{align*}
	|\braket{\psi_{f}^A}{\psi_{f}^B}|
    &\leq |\braket{v_A}{\psi_{f}^B}|+ |\braket{u_A}{\psi_{f}^B}|
    \leq |\braket{v_A}{\psi_{f}^B}|+ \|u_A\|\cdot\|\psi_{f}^B\|
    \leq |\braket{v_A}{\psi_{f}^B}|+ 0.1\\
    &\leq |\braket{v_A}{v_B}|+|\braket{v_A}{u_B}|+0.1
    \leq |\braket{v_A}{v_B}|+\|v_A\|\|u_B\|+0.1
    \leq |\braket{v_A}{v_B}|+0.2 \\
    &= 0.2+\left|\sum_{(a,b)\in \Lambda(A)\cap\Lambda(B)}(\beta^A_{ab})^*\beta^B_{ab}\langle\phi^A_{ab}|\phi^B_{ab}\rangle\right|
	\leq 0.2 + \sum_{(a,b)\in \Lambda(A)\cap\Lambda(B)}|\beta_{ab}^A||\beta_{ab}^B|.
    \end{align*}
	
	\allowdisplaybreaks
	
	Let $\rchi_{\Lambda(A)}$ be defined by
	$\rchi_{\Lambda(A)}(a,b)=1$ if $(a,b)\in \Lambda(A)$
	and $\rchi_{\Lambda(A)}(a,b)=0$ otherwise. We have 
	\begin{align*}
	&\E_{(A,B)\sim\mathcal{R}}
	\left(|\langle\psi_f^A|\psi_f^B\rangle|\right)-0.2
	\leq\E_{(A,B)\sim\mathcal{R}}\left[
	\sum_{(a,b)\in \Lambda(A)\cap \Lambda(B)}
	|\beta_{ab}^A||\beta_{ab}^B|\right] & \\
	&=\frac{1}{|\mathcal{R}|}\sum_{(A,B)\in\mathcal{R}}\sum_{(a,b)\in \Lambda(A)\cap \Lambda(B)}
	|\beta_{ab}^A||\beta_{ab}^B| &\\
	&\leq\frac{1}{|\mathcal{R}|}\sqrt{
		\sum_{(A,B)\in\mathcal{R}}
		\sum_{(a,b)\in \Lambda(A)\cap \Lambda(B)}
		|\beta_{ab}^A|^2
		\sum_{(A,B)\in\mathcal{R}}
		\sum_{(a,b)\in \Lambda(A)\cap \Lambda(B)}
		|\beta_{ab}^B|^2
	} &\mbox{(Cauchy-Schwartz)}\\
	&=\frac{1}{|\mathcal{R}|}
	\sum_{(A,B)\in\mathcal{R}}
	\sum_{(a,b)\in \Lambda(A)\cap \Lambda(B)}
	|\beta_{ab}^A|^2 &
	\mbox{(since }\mathcal{R}\mbox{ is symmetric)}\\
	&=\E_{(A,B)\sim\mathcal{R}}\left[
	\sum_{(a,b)\in \Lambda(A)\cap \Lambda(B)}
	|\beta_{ab}^A|^2\right] & \\
	&=\E_{A\sim S(n)}\left[\E_{B\sim\mathcal{R}_A}\left[
	\sum_{(a,b)\in \Lambda(A)\cap \Lambda(B)}
	|\beta_{ab}^A|^2\right]\right] & \\
	&=\E_{A\sim S(n)}\left[\E_{B\sim\mathcal{R}_A}\left[
	\sum_{(a,b)\in \Lambda(A)}
	|\beta_{ab}^A|^2\rchi_{\Lambda(B)}(a,b)
	\right]\right] & \\
	&=\E_{A\sim S(n)}\left[\sum_{(a,b)\in \Lambda(A)}
	|\beta_{ab}^A|^2
	\E_{B\sim\mathcal{R}_A}\left[\rchi_{\Lambda(B)}(a,b)
	\right]\right] &\mbox{(linearity of expectation)} \\
	&\leq\E_{A\sim S(n)}\left[\sum_{(a,b) \in \Lambda(A)}
	|\beta_{ab}^A|^2\right]
	\max_{\substack{A\in S(n), (a,b)\in \Lambda(A)}}\left(
	\E_{B\sim\mathcal{R}_A}[\rchi_{\Lambda(B)}(a,b)]
	\right) & \\
	&\leq\max_{\substack{A\in S(n), (a,b)\in \Lambda(A)}}\left(
	\E_{B\sim\mathcal{R}_A}[\rchi_{\Lambda(B)}(a,b)]
	\right) & \\
	&=\max_{A\in S(n),(a,b)\in \Lambda(A)} \Pr_{B \sim \mathcal{R}_{A}}[(a,b) \in \Lambda(B)]. & \qedhere
	\end{align*}
\end{proof}

It remains to bound this expression.

\begin{lemma} \label{lem:probability_a_in_B_upperbound}
	\[ \max_{A \in S(n), (a,b)\in \Lambda(A)}
	\Pr_{B \sim \mathcal{R}_{A}}[(a,b)\in \Lambda(B)] \leq \frac{1}{4}.\]
\end{lemma}

\begin{proof}
	Fix $A\in S(n)$ and $(a,b)\in \Lambda(A)$,
	so $a\neq b$, $a\in A$, $b\in A^\perp$, and $a,b\neq 0$.
	Picking $B\sim\mathcal{R}_A$ is picking a vector
	space whose intersection with $A$ has dimension $n/2-1$.
	One way of picking such a vector space uniformly
	at random is by picking a random basis for $A$,
	then discarding one of the vectors in the basis
	and adding a vector outside of $A$. This gives
	a new set of $n/2$ independent vectors, which
	we take to be a basis for $B$.
	
	Let $G(m,k)$ be the number of subspaces of
	$\mathbb{F}_2^m$ of dimension $k$.
	We can write down an expression for $G(m,k)$
	using the following argument:
	there are $(2^m-1)(2^m-2)\ldots(2^m-2^{k-1})$
	ways to pick an ordered list of $k$ independent vectors
	in $\mathbb{F}_2^m$, and each such list corresponds
	to a subspace. However, this over-counts the subspaces
	by a factor of $(2^k-1)(2^k-2)\ldots(2^k-2^{k-1})$.
	We conclude that\footnote{In more generality, the number of subspaces of dimension $k$ in a vector space of dimension $m$ over a field with $q$ elements is the Gaussian binomial coefficient~\cite{Prasad2010}, $\binom{m}{n}_{q}= \prod_{i=0}^{k-1}\frac{q^{m-i}-1}{q^{k-i}-1}$.}
	\[G(m,k)=\prod_{i=0}^{k-1}\frac{2^{m-i}-1}{2^{k-i}-1}.\]
	
	We are interested in the probability that $(a,b)\in \Lambda(B)$.
	This is simply the probability that $a\in B$ and $b\in B^\perp$.
	$a$ is in $B$ if $a$ is in the span
	of the $n/2-1$ vectors we chose from $A$, and
	$b$ is in $B^\perp$ if $b$ is orthogonal to the vector
	we chose from outside of $A$. It is clear that
	these events are independent.
	
	The former happens with probability
	\[ G(n/2-1,n/2-2)/G(n/2,n/2-1)
	=\frac{1}{2}\left(1-\frac{1}{2^{n/2}-1}\right)\leq 1/2.\]
	For the latter, note that exactly half the vectors
	in $\mathbb{F}_2^n$ are orthogonal to $b$,
	and $2^{n/2}$ of those are in $A$. So the probability
	of picking an orthogonal vector to $c$ outside of $A$ is
	\[\frac{2^{n-1}-2^{n/2}}{2^n-2^{n/2}}
	=\frac{2^{n/2-1}-1}{2^{n/2}-1}\leq 1/2.\]
	Thus, the probability of both events occurring is at most $1/4$.
\end{proof}

%

From these lemmas, we can use 
Theorem~\ref{thm:inner-product-adversary-method}
to get a lower bound on the number of queries which
produces a pair $(a,b)\in\Lambda(A)$ with probability
at least $0.99$.

\begin{theorem}	\label{thm:bounded-error-counterfeiting}
	Any algorithm which takes $\ket{A}$ as input
	and outputs (with success probability at least $0.99$)
	a pair $(a,b)\in \Lambda(A)$ for all subspaces $A \in S(n)$, must make
	$\Omega( 2^{n/4})$ oracle queries.
\end{theorem}

\begin{proof}
	The algorithm starts with the initial state
	$\ket{\psi^A_{init}}=\ket{0,0}\ket{0}_W\ket{A}\ket{0}$
	and ends with $\ket{\psi^A_{f}}$. Note that if $(A,B)\in\mathcal{R}$, then
	$|\braket{\psi^A_{init}}{\psi^B_{init}}|=|\braket{A}{B}|=1/2$.
	Also, by Lemma \ref{lem:upper-bound-on-final-states} and Lemma~\ref{lem:probability_a_in_B_upperbound}, we have
	$\E_{(A,B) \sim \mathcal{R}}|\braket{\psi_{f}^{A}}{\psi_{f}^{B}}| \leq \frac{1}{4}+0.2=0.45.$
	Therefore, by Theorem~\ref{thm:inner-product-adversary-method},
	the algorithm must make $\Omega\left( 2^{n/4}\right) $ oracle queries. 
\end{proof}

\subsection{Strengthening the Lower Bound}

We've shown that there is no way to find such pairs, with probability 
at least $0.99$, for all $A$s. We will next show that there is no way 
to find the pairs even with exponentially small success probability, 
and even for a random $A$. We do that in two steps. The first step 
reduces exponentially small probability
to constant probability by providing an amplification
procedure,
and the second step reduces average $A$ to worst case $A$.

For the first step, we will need the following theorem from \cite{aaronson13quantum}.

\begin{theorem}[{\cite[Theorem 8]{aaronson2012quantum}}]
	\label{thm:fixed_point_search}
	Let $U_{Init}= I - 2\ketbra{Init}$, and for a subspace $G$,
	$U_{G} \ket{v} = -\ket{v}$ for all $\ket{v} \in G$
	and acts as the identity on all $\ket{v}$ orthogonal to $G$.
	Let $\delta \geq 2 \epsilon>0$.
	Suppose $F(\ket{Init},G) \geq \epsilon$.
	There exists an algorithm which starts with the state $\ket{Init}$, uses $O(\frac{\log 1/\delta}{\epsilon \delta^{2}})$ oracle calls to $U_{Init}$ and $U_{G}$, and prepares a state $\rho$ such that $F(\rho,G) \geq 1 -\delta$.
\end{theorem}

Here\footnote{The fidelity between a mixed state and a pure state is $F(\rho,\ket{\psi})=\sqrt{\bra{\psi}\rho\ket{\psi}}$. The above definition is a natural extension to the distance between a mixed state and a subspace.} $F(\rho,G) = \max_{\ket{\psi} \in G} \sqrt{\bra{\psi} \rho \ket{\psi}}$. Note that $F(\rho,G)^{2}=\max_{\ket{\psi} \in G} \tr(\ketbra{\psi} \rho) \leq \tr(\Pi_{G} \rho)$. 

Equipped with this tool, we now reduce security against
exponentially small success probability to security against constant
success probability.

\begin{corollary}
	Let $1/\epsilon=o\left(2^{n/2}\right)$. Given one copy of $\ket{A}$ and oracle access to $U_{A^{*}}$, a counterfeiter needs $\Omega\left(\sqrt{\epsilon} 2^{n/4}\right)$ queries to prepare a state $\rho$ such that $\tr(\Pi_{C(A)}\rho) \geq \epsilon$ for all $A \in S(n)$.
	\label{cor:high_error_worst_case}
\end{corollary}
\begin{proof}
	Suppose we have a counterfeiter $C$ that, on input $\ket{A}$,
	makes $T=o\left(\sqrt{\epsilon} 2^{n/4}\right)$ queries
	and produces a pair $(a,b)\in\Lambda(A)$ with probability at least $\epsilon$.
	We will construct another counterfeiter $C'$ which violates Theorem~\ref{thm:bounded-error-counterfeiting} by producing a pair
	$(a,b)\in\Lambda(A)$ with probability $0.99$.
	
	If the success probability was exactly $\epsilon$, we could have used amplitude amplification. The problem, which is the same problem Aaronson and Christiano had to deal with, is the so called souffl\'e problem ~\cite{brassard07searching} of  amplitude amplification (and also Grover search): the  algorithm must run (the souffl\'e must be in the oven) just for the right time.
	This can be easily circumvented if $\ket{Init}$ can be prepared free of charge -- which is \emph{not} our case. Algorithms which do not have this property are called \emph{fixed point quantum search}. Previously, the fixed point quantum search algorithms~\cite{tulsi06new,chakraborty05bounds} lost their quantum speedup, though they had an exponential convergence.  Aaronson and Christiano constructed an algorithm which has the quadratic speedup but loses the exponential convergence.
	
	Specifically, we apply Theorem~\ref{thm:fixed_point_search} with
	$\ket{init}=\ket{\psi_f^A}$, the final state of the counterfeiter $C$ when
	run on $A$, and $G$ being the subspace spanned by
	the computational basis states of the algorithm that contain outputs in $\Lambda(A)$: that is, states of the form $\ket{c,p,x}\ket{s}_W\ket{a}\ket{b}$
	with $(a,b)\in\Lambda(A)$.
	We note that $U_{G}$ can be implemented using two queries to  $U_{A^{*}}$.
	Implementing the operator $I - 2\ketbra{A}$ can be done using one query to $U_{A}$ and one query $U_{A^{\perp}}$, as shown by
	Aaronson and Christiano (more details can also be found in Section~\ref{sec:AC_overview}).
	Since $\ket{\psi_{init}^A}$ contains only the state $\ket{A}$ plus
	some blank registers, we can also implement the operator
	$I-2\ketbra{\psi_{init}^A}$ in this way.
	Therefore $U_{Init}$ can be implemented using one call for $C$, one call for $C^{\dagger}$, and two calls for $U_{A^{*}}$:
	\[ U_{Init}=I-2\ketbra{\psi_f^A}= I-2C\ketbra{\psi_{init}^A}C^{\dagger}
	= C(I-2\ketbra{\psi_{init}^A})C^{\dagger}.\]
	
	Note that the success probability of $C$ is equal to
	$F(\ket{\psi_f^A},G)^2$. Since this probability
	is at least $\epsilon$, we get $F(\ket{\psi_f^A},G)\geq \sqrt{\epsilon}$.
	We apply \ref{thm:fixed_point_search} with $\delta=0.005$ and
	$\epsilon$ (in the theorem) set to $\sqrt{\epsilon}$ (we
	can assume without loss of generality that $\sqrt{\epsilon}\leq 0.0025$).
	Theorem~\ref{thm:fixed_point_search} gives us an algorithm
	$C'$ which outputs a state $\rho$ with $F(\rho,G)\geq 0.99$, and which uses $O(\frac{1}{\sqrt{\epsilon}})$ calls to the counterfeiter $C$ and $U_{A^{*}}$.
	Note that $0.99\leq 0.995^2\leq F(\rho,G)^2\leq \tr(\Pi_G\rho)$,
	which means that measuring $\rho$ provides a pair $(a,b)\in\Lambda(A)$
	with probability at least $0.99$.
	Moreover, since each call for $C$ uses $T$ queries to $U_{A^{*}}$, the total number of queries used by $C'$ is
	\[ O\left(\frac{1}{\sqrt{\epsilon}}\right) \cdot (1 + T)
	=O\left(\frac{1}{\sqrt{\epsilon}}\right)+O\left(\frac{T}{\sqrt{\epsilon}}\right).\]
	However, $C'$ must use $\Omega(2^{n/4})$ queries
	by Theorem~\ref{thm:bounded-error-counterfeiting}.
	Since $1/\sqrt{\epsilon}=o(2^{n/4})$, it follows that
	$T/\sqrt{\epsilon}=\Omega(2^{n/4})$, or $T=\Omega(\sqrt{\epsilon}2^{n/4})$,
	as desired.
\end{proof}

We now complete the proof of Theorem~\ref{thm:lower-bound-average-case},
which we restate here for convenience.

\LowerBoundAverageCase*

\begin{proof}
	The only missing ingredient is a worst case to average case reduction.
	Suppose we had a counterfeiter $C$ which violated the above. We will construct another counterfeiter $C'$ which violates Corollary~\ref{cor:high_error_worst_case}. $C'$ first chooses a random linear invertible map $f:\mathbb{F}_{2}^{n} \rightarrow \mathbb{F}_{2}^{n}$. Let $f(A)$ be the image of $A$ under $f$. For any fixed $A$, $f(A)$ is a uniformly random element of $S$.  Let $U_{f},U_{f^{-1}}$ be the unitary maps $U_{f}\ket{x}= \ket{f(x)},\ U_{f^{-1}}\ket{x}= \ket{f^{-1}(x)}$. Since $U_{f(A)^{*}}=U_{f}U_{A^{*}} U_{f}^{\dagger}$, the oracle $U_{f(A)^{*}}$ can be efficiently implemented using 1 query of $U_{A^{*}}$, and $\ket{f(A)}$ can be prepared using $\ket{A}$. It then applies the counterfeiting procedure to $\ket{f(A)}$, and since $f(A)$ is random in $S(n)$, it finds $(a,b)\in \Lambda(f(A))$  using $t$ queries with error probability $\epsilon$. Using $f^{-1}$, it can find $(f^{-1}(a),f^{-1}(b)) \in \Lambda(A)$, which contradicts Corollary~\ref{cor:high_error_worst_case}.
\end{proof}
 
\section{The One-time Scheme}
\label{sec:one-time_scheme}
Theorem~\ref{thm:lower-bound-average-case} can be used to show that if we treat the membership to $A$ and $A^\perp$ as an oracle, we can construct a one-time secure TS scheme. The challenge that we are facing was described by Aaronson and Christiano as follows (note that their context was public quantum money, hence the reference to a \emph{bank}):
\begin{displayquote}
Given a subspace $A \preceq \mathbb F_{2}^{n}$, how can a bank distribute an ``obfuscated program'' $\mathcal O(\IAH)$, which legitimate buyers and sellers can use to decide membership in both $A$ and $A^{\perp}$, but which does not reveal anything else about A that might facilitate counterfeiting?
\end{displayquote}

For a gentle bird's-eye view on program obfuscation, see Barak's excellent, though already somewhat outdated, review~\cite{barak16hopes}.
There are several notions of program obfuscation in the literature. In all of them, an obfuscator is a (probabilistic) compiler which takes a circuit \nom{$C$}{A circuit} and outputs a classical obfuscated program
\nom{$\mathcal O(C)$}{An obfuscated program of the circuit $C$} which preserves the functionality, i.e.,
\begin{equation}
\forall x,\ C(x)=\mathcal O(C)(x),
  \label{eq:obfuscation_functionality}
\end{equation}
and $\mathcal O(C)$ is ``unintelligible'' in some way. The notion which we use for  unintelligible is called Virtual Black Box (\nom{VBB}{virtual black box, a form of obfuscation}), which requires that for every polynomial adversary with access to $\mathcal O(C)$ there exists an efficient simulator with only oracle access to $C$, which outputs a computationally indistinguishable distribution with respect to the adversary's distribution. Barak et al.~ \cite{barak12on} defined this notion and showed that there is a class of circuits for which VBB obfuscation is impossible (hence \emph{general purpose} VBB obfuscation is impossible).  As far as we know, their results, as well as the other negative results related to VBB (e.g. ~\cite{goldwasser05impossibility,bitansky14impossibility}), does not rule out specific black box obfuscation for the class of functions required in this work.

Let $\mathfrak{S}$ be a set of circuits that compute $\IAH$ for all $A \in S(n)$. 
On a positive note for our purposes, Canetti, Rothblum and Varia~\cite{canetti10obfuscation} proved VBB obfuscation for hyperplane membership of $\mathbb{F}_{p}^{n}$ but only for constant $n$ and a large $p$. Their construction is not secure against quantum adversaries, therefore we do not attempt to adapt our scheme to fit to their setting. 




Next we define VBB obfuscation which will be sufficient for our construction.

\begin{definition}[VBB  Obfuscator]
 A probabilistic algorithm $\mathcal O$ is a VBB circuit obfuscator with dependent auxiliary input
for the collection $\mathfrak F$ of circuits if the following three conditions hold:
\begin{enumerate}
\item{(Functionality.)} For every circuit $C \in \mathfrak F$, the string $\mathcal O(C)$ is a description of a classical circuit that computes the same function as $C$.

\item{(Virtual black box).}
For any QPT $\Adv$ there is a QPT $\Sim$ such that  for every $C \in \mathfrak{F}$ and auxiliary state $\ket{\psi}$, the distributions $\Adv(\mathcal O(C),\ket{\psi})$ and $\Sim^{C}(1^{|C|},\ket{\psi})$ are computationally indistinguishable.

For our purposes, we only need classical computational indistinguishability, i.e., for every PPT distinguisher $\Dist$,  
\[ |\Pr\left[\Dist(\Adv(\mathcal O(C),\ket{\psi}))=1\right]- \Pr\left[\Dist(\Sim^{C}(1^{|C|},\ket{\psi}))=1\right] |\leq \negl(|C|). \]

\item{(Efficiency.)} The algorithm $\mathcal O$ is PPT.\footnote{Note that efficiency also implies that  the slowdown by running $\mathcal O(C)$ instead of $C$  is at most polynomial.}
\end{enumerate}
\label{def:vbb_requirements}
\end{definition}

 This definition differs from the  VBB property given in Barak et al.\cite{barak12on} in several ways:
\begin{itemize}
\item We need security against a QPT adversary, while theirs is against a PPT adversary.
\item We need security even when the adversary has access to auxiliary input states, and furthermore, that auxiliary input state depends on the circuit $C$, see \cite{goldwasser05impossibility,bitansky14impossibility}. We need security with \emph{dependent} auxiliary input because the adversary have a quantum singing token, which depends on the circuit.  
\item We require computational indistinguishability between the distributions that the adversary and the simulator generate. The main impossibility results in~\cite{barak12on} assumed a weaker notion of security, in which the output of the adversary is a bit (thus, the goal of the simulator is learning a predicate). 
\end{itemize}

We note that our security proof holds even if the VBB property holds only for a random circuit in $\mathfrak{F}$.

We are now ready to describe our scheme in Algorithm~\ref{alg:ot1}, denoted $\ot1$, which is a onetime $1$-restricted, testable TS scheme. As we have seen in Section~\ref{sec:standard-construction}, this can be mapped to a full blown (without the onetime and length restrictions) testable TS scheme. 

\begin{algorithm}
  \caption{$\ot1$ scheme.}
   
    \begin{algorithmic}[1] 
        \Procedure{$\keygen$}{$1^{\kappa}$}
            \State  Set $n=f(\kappa)$ \Comment{$f$ can be any super-logarithmic function.} 
            \State $A\getsr S(n)$
            \State $pk \gets \mathcal O(\IAH)$  \Comment{See Eq.~\eqref{eq:IAH} for the definition of $\IAH$, and Definition~\ref{def:vbb_requirements} for $\mathcal O$.} 
\label{line:obfuscation}
            \State $sk \gets \langle A\rangle $ \Comment{$\langle A \rangle$ is a (classical) basis for $A$. }
            \State \textbf{return} $(pk,sk)$
        \EndProcedure

        \Procedure{$\tokengen$}{$\langle A \rangle$}
            \State $\ket{\stamp} \gets \ket{A}$ \Comment{Given a basis for $A$, the state $\ket{A}$ can be generated efficiently~\cite{aaronson2012quantum}.}
            \State \textbf{return} $\ket{\stamp}$
        \EndProcedure

       \Procedure{$\sign$}{$\alpha \in \{0,1\},\ket{A}$}
            \State Apply $H^{\otimes n}$ iff $\alpha=1$ to the state $\ket{A}$. 
 \Comment{$H^{\tensor n}\ket{A}=\ket{A^{\perp}}$}
            \State Measure the resulting state  in the standard basis. Set $sig$ to be the outcome. \label{line:sign_measurement}
            \If{$sig=0 \ldots 0$}
                \State \textbf{return} failed
            \Else
                \State \textbf{return} $sig$
            \EndIf
        \EndProcedure

        \Procedure{$\verify_{pk}$}{$\alpha \in \{0,1\}, sig$}            
            \If{$sig=0^n$}
                \State \textbf{return} $0$
            \EndIf
            \State \textbf{return} $\mathcal O (\IAH)(\alpha,sig)$.
        \EndProcedure

        \Procedure{$\verifytoken_{pk}$}{$\tau$}            
            \State Using $\mathcal O(\IAH)$, measure the state $\tau$ using the two outcome measurement $\{\ketbra{A}, I-\ketbra{A}\}$, following the procedure in  Eq.~\eqref{eq:verification}. If the outcome is the first, \textbf{return} $(T,\ket{A})$. Otherwise, \textbf{return} $F$. 
        \EndProcedure
 
    \end{algorithmic}
\label{alg:ot1}
\end{algorithm}

{}

\begin{remark} The procedures $\keygen$ and $\verify$ are entirely classical. The state $\ket{A}$ generated in $\tokengen$ can be constructed using only Clifford gates~\cite{cleve97efficient,grassl03efficient}, where the total number of gates is polynomial in $n$ and hence poly-logarithmic in $\kappa$. Clifford gates do not form a universal quantum gate set, and may be easier to implement on certain architectures, see, e.g.~\cite{bombin11clifford}.  
\end{remark}
\begin{theorem}\label{thm:vbb-obfuscation}
If $\mathcal O$ is a VBB obfuscator with dependent auxiliary input for the class $\mathfrak S$, then the scheme $\ot1$ described in Algorithm~\ref{alg:ot1} is an imperfect onetime everlasting secure, 1-restricted, testable,
unforgeable tokenized signature scheme.
\end{theorem}

\begin{proof}
The imperfect completeness property (see Eq.~\ref{eq:imperfect_ts}) for 1-restricted scheme is justified by the construction. The scheme is imperfect (and not perfect) since with probability of $\frac{1}{2^{n/2}}=\negl(\kappa)$ the outcome of the measurement in line ~\ref{line:sign_measurement} is $0 \ldots 0$ (which causes $\sign$ to fail). This is the only element that is trivial to guess in $A$ and $A^\perp$, since the $0\ldots 0$ vector is an element of every subspace of $\mathbb F_{2}^{n}$. 

We show that the scheme is unforgeable. Fix $A$. Suppose the distinguisher 
runs the (classical) algorithm $\verify_{2, pk(A)}$, 
and the auxiliary state is $\ket{A}$. By the VBB property\footnote{Here we took advantage of the fact  that $\verify$ in our scheme is classical. This should clarify why our requirement in the VBB property is classical computational indistinguishability (and not quantum indistinguishability).}, there exists a QPT $\Sim$ such that
\[ |\Pr_{\Adv, \mathcal O}[\verify_{2,pk(A)}(\Adv(\mathcal O(\IAH),\ket{A}))=1]- \Pr_{\Sim}[\verify_{2,pk(A)}(\Sim^{\IAH}(1^{|\IAH|},\ket{A}))=1] |\leq \negl(\kappa). \] 

The above holds for every $A \in S$. We now take the average over all $A \in S$:
\[ |\Pr_{A \getsr S, \Adv, \mathcal O}[\verify_{2,pk(A)}(\Adv(\mathcal O(\IAH),\ket{A}))=1]- \Pr_{A \in_{R} S,\Sim}[\verify_{2,pk(A)}(\Sim^{\IAH}(1^{|\IAH|},\ket{A}))=1] |\leq \negl(\kappa) \] 
By Theorem~\ref{thm:lower-bound-average-case}, the second term of the above equation can be bounded. For every QPT $\Sim$,
\begin{equation}
\Pr_{A \getsr S,\Sim}[\verify_{2,pk(A)}(\Sim^{\IAH}(1^{|\IAH|},\ket{A}))=1]=\frac{\poly(\kappa)}{2^{n/2}}=\negl(\kappa).
\end{equation}
Combining the above two equations, we conclude that onetime super-security (see Definition~\ref{def:super_security}) holds:
\begin{equation}
\Pr_{A\getsr S, \Adv, \mathcal O}[\verify_{2,pk(A)}(\Adv(\mathcal O(\IAH),\ket{A}))=1]\leq \negl(\kappa).
\end{equation}

The testability properties (see Definition~\ref{def:testability}) also follow trivially by construction because $\verifytoken$ implements the rank-1 projector of the token $\ket{\stamp}=\ket{A}$ -- see Eq.~\eqref{eq:verification}.

The everlasting revocability is also simple to prove. All the adversary has is one copy of the state $\ket{A}$ (and no access whatsoever to $\IAH$ since in the everlasting revocability definition, the adversary is not given the public key). Theorem~\ref{thm:lower-bound-average-case} shows that in order to find one non-trivial element from $A$ and another from $A^{\perp}$, exponentially many oracle queries to $\IAH$ are required in order to succeed with non-negligible probability. In particular, this task cannot be achieved with no queries to the oracle. 
\end{proof}

Constructing a dependent-auxiliary-input
VBB obfuscator for $\mathfrak{S}$ based on some
reasonable cryptographic primitive remains an interesting open
problem and therefore our construction cannot be instantiated. In previous versions of this work, we conjectured that replacing the VBB obfuscation with indistinguishability obfuscation would also be secure. This turned up to be correct for quantum money --- see~\cite{Zha21}. It is still unclear whether this approach works for tokens for digital signatures. Recently, Coladangelo et al.~\cite{CLLZ21} proved that a variant of the approach above, which uses coset states (instead of subspace states), and uses indistinguishability obfuscation to generate the public key, is indeed secure.

\section{Applications}
\label{sec:applications}
\subsection{Sending Quantum Money over a Classical Channel}
\label{sec:mult-bank-branch}

In this section, we describe several interesting properties of
a quantum money scheme that is derived from a tokenized 
(either private or public) signature scheme. In particular, we show how to convert a quantum
bill into a classical ``check'', which is addressed to a specific person
and can be exchanged for a quantum bill at a bank branch.

Recall from Theorem~\ref{thm:testable_ts_is_qm}, and its private analog (see Appendix~\ref{sec:MAC}), that a testable public (private)
digital signature scheme satisfies the definition of a public (resp. private)
quantum money scheme, with the signing tokens acting as
the quantum bills. Now, in this setting, if Alice holds a quantum bill,
one thing she can do is spend it the usual way. However, an alternative thing she can do with
the bill is use it to sign a message. Such a signature will necessarily
consume the bill, and hence can be used as proof that Alice has burned
her bill.

In particular, consider what happens if Alice signs the message,
``I wish to give a bill to Bob''. Alice can send over this classical signature
to Bob, who can verify the signature using the public verification scheme.
Bob can then show this signature to his bank branch. Crucially, the bank
knows that Alice burned her bill, since the signature is valid; hence
the bank can safely issue Bob a bill. In essence, this usage
converts a quantum bill into a classical check!

There are a few potential flaws with this idea, particularly relating to double
spending: what happens if Bob tries to cash his check twice?
To prevent this, Bob can ask Alice to add a serial number or a time stamp to the check, signing
a message like ``I wish to give a bill to Bob; the current date is September
27, 2016, at 11:59pm''. The time stamp can be verified by Bob to be
accurate, and ensures that Alice does not give Bob the same check twice.
The bank can then keep a database of all the checks cashed,
and reject duplicate cashing of the same check.

This solution requires the bank to keep a database of cashed checks,
which raises a further issue: what if Bob tries to cash the check
at two different branches simultaneously, without the branches
getting a chance to compare notes? To address this issue,
we add a further modification to the scheme: Bob can tell Alice
the name of the branch he plans to go to (e.g. ``branch number 317''),
and Alice can add this branch name to her check
(``this check cashable at branch number 317''). The bank branch
can then only accept checks that specify that branch in this way.
This allows each branch to keep a completely separate database,
without any need for communication between the branches. 


The scheme above has several useful properties, which are listed below. The first two are relevant to both variants (public and private), and the last two are relevant only for public tokenized signature scheme.

\begin{enumerate}
	\item If Alice loses quantum communication with the outside world,
	she can still use her money for payment -- even without the entanglement
	that would be needed for quantum teleportation. The only schemes which were shown to be classically verifiable are~\cite{gavinsky2012quantum,georgiou15new}. Our scheme achieves that with only one round of communication. Therefore, We call this a ``check": verification is not of the type of a challenge and response, like the previously classically verifiable quantum money schemes mentioned before.
	\item The bank branches in this scheme do not need to communicate
	in order to cash checks; everything can be done completely offline.
        Furthermore, each branch's data remain static -- no data needs to be updated, or even be stored (besides the scheme's key)~\cite{gavinsky2012quantum}. This can be done if, for example, each bank branch
     only accepts checks once every minute, and only if the check's timestamp
     falls in that minute. In this case the branch does not need to store
     a database of previous checks cashed.
	\item Public quantum money has the nice property that the transactions
	are \emph{untraceable}, so that third parties need not know that
	the transaction took place. While this is an advantage for privacy,
	it can also be a disadvantage: if Alice pays Bob, he can deny receiving
	the money, and Alice has no way to prove that she has paid. With our
	scheme, Alice can \emph{choose} the method of payment:
	either the untraceable way, or else by using the traceable classical check,
	which allows Alice to prove that she has paid. Proving the payment, say, to the police, does not require the  bank's cooperation. Additional advantages, such as logging, auditing, and the ability to recover from communication errors were discussed in~\cite{RS20}.
	
    \item In the public setting, the bank branches need not store the secret key of the quantum
	money scheme. This means that if a branch is robbed or hacked,
	the thief gains the quantum bills stored in the branch,
	but does not gain the ability to mint new bills; no long-term damage
	to the monetary system is inflicted. None of the private money schemes have this property. 
\end{enumerate}

The main disadvantages of the scheme are:
\begin{itemize}
\item Alice needs the ability to perform universal quantum computation
and store \emph{entangled} quantum states long-term, unlike several private money schemes~\cite{wiesner1983conjugate,bennett1983quantum,gavinsky2012quantum,georgiou15new}.
\item Our public scheme requires computational assumptions, and even worse, non-standard ones. The private setting is comparable to the previous constructions in terms of computational assumptions: if one does not care that the key length grows linearly with the number of money states in the system, no computational assumptions are needed; otherwise, only a  weak assumption, namely, a quantum secure one way function is required. 
\item Our scheme does not
tolerate constant level of noise out of the box~\cite{Pastawski02102012}.
\end{itemize}

\subsection{Circumventing Two Faced Behavior}
\label{sec:two_faced}
Consider a typical scenario in distributed computing, which involves Alice, Bob and Charlie (although the following applies when there are $n$ parties). In an honest scenario, Alice has some preference,  which she is supposed to send to Bob and Charlie. But Alice, as well as Bob and Charlie may be malicious. In consensus problems, it is crucial that honest Bob and Charlie will output the same outcome, even if Alice is malicious. One barrier is if Alice pretends that her preference is $x$ to Bob, and $y \neq x$ to Charlie: this is called \emph{a two-faced behavior}. If Charlie tells Bob that Alice's preference is $x$, and Alice tells Bob that her  preference is $y \neq x $, Bob cannot determine whether Alice or Charlie (or both) is the cheater. 

If we assume that Alice receives exactly one signing token, the two-faced behavior can be circumvented: she can only sign one message, and therefore cannot send two contradicting signed messages to Bob and Charlie. This should be compared to the more complicated solution, which only assumes digital signatures (see ~\cite[Section 6.2.4]{lynch96distributed} and references therein): 1) Alice sends her signed preference to Bob and Charlie. 2) Bob and Charlie compare the signed messages they received. If the preferences are not the same, and both pass the verification, they conclude that Alice is the cheater. 

Even though the construction and analysis of the solution which uses signing tokens is simpler than the one which uses digital signatures, as far as we know, it does not improve the asymptotic running time for problems such as the Byzantine agreement and its variants.
\section{Open Questions}
Are there any other tokenized schemes? The immediate candidates are variants of digital signatures, such as blind signatures, group signatures and ring signatures. Can the techniques that were introduced in this work be used to construct quantum copy-protection~\cite{aaronson2009quantum}?

Can we add the testability property to any TS scheme, similarly to the way we showed how revocability can be added (see Theorem~\ref{thm:revocability})? We can show that there is a test which guarantees that the signing token can be used to sign a random document w.h.p.; but we cannot construct a test which guarantees that the signing token can be used to sign every document w.h.p.

\section{Acknowledgments}

We thank Scott Aaronson, Dorit Aharonov, Nir Bitansky, Zvika Brakerski, Aram Harrow, 
 Robin Kothari, Gil Segev and Vinod Vaikuntanathan for valuable discussions. We thank Paul Christiano for reporting to us the attack described in Section~\ref{sec:attack-noisy-scheme}, and for his other comments. We also thank the anonymous referees for their comments.

Most of this work was done while S.B.D. was at MIT and O.S. was at the Hebrew University and MIT.

This work was partially supported by NSF grant no. 2745557, and ERC Grant 030-8301, the Israel Science Foundation (ISF) grant No. 682/18 and 2137/19, and
by the Cyber Security Research Center at Ben-Gurion University. The authors acknowledge the hospitality of the Center for Theoretical Physics at MIT. The icon $\stamp$ was downloaded from \url{http://icons8.com}, and is licensed under Creative Commons Attribution-NoDerivs 3.0 Unported.

\BeforeBeginEnvironment{wrapfigure}{\setlength{\intextsep}{0pt}}
\begin{wrapfigure}{r}{90px}
    \includegraphics[width=40px]{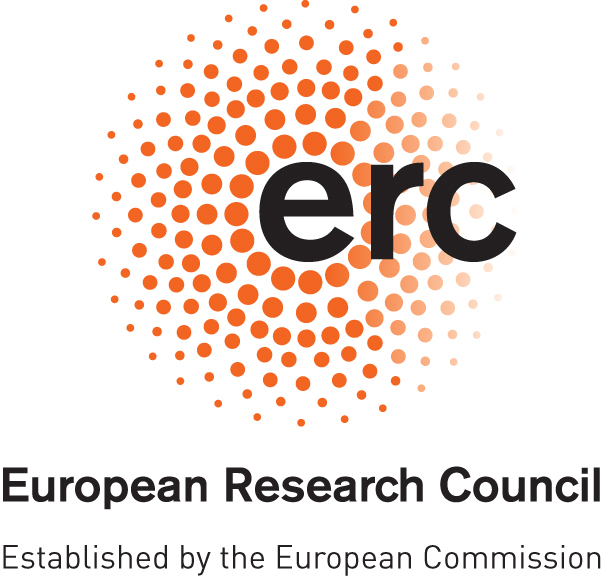}  \includegraphics[width=40px]{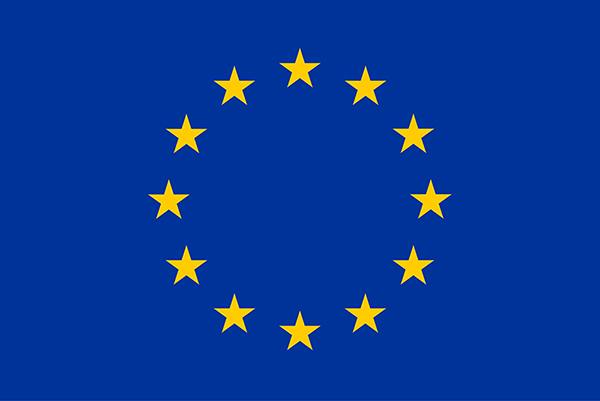}
\end{wrapfigure}This work has received funding from the European Research Council (ERC) under the European Union’s Horizon 2020 research and innovation programme (grant agreement No 756482).

\bibliography{tokenized}
\bibliographystyle{alphaabbrurldoieprint}

\appendix
\section{Nomenclature}
\label{sec:nomenclature}

\begin{nopreview}
\printnomenclature
\end{nopreview}

\section{An Attack on Aaronson-Christiano's Scheme}
\label{sec:attack_fix_quantum_money}
\subsection{Overview}
\label{sec:AC_overview}
We start with an overview of Aaronson-Christiano's scheme~\cite{aaronson13quantum}. The quantum money state issued by the bank is $\ket{A}$ (see Eq.~\eqref{eq:ketA}) for a random $A\in S(n)$, together with a classical string which is used for verifying that state -- additional details on this classical string will be given shortly.
They define \nom{$\PA$}{The projection to the span of the elements in $A$} and $\PAD$ as the projection to the span of $A$ and $A^\perp$ respectively,
\begin{equation}
\PA\equiv\sum_{a\in A} \ketbra{a},\ \PAD\equiv \sum_{a\in A^\perp} \ketbra{a}
\label{eq:PA}
\end{equation}
and show that
\begin{equation}
H^{\otimes n}\PAD H^{\otimes n}\PA= \ketbra{A}.
\label{eq:verification}
\end{equation}
Eq.~\eqref{eq:verification} implies that the state $\ket{A}$ can be verified efficiently using 2 queries to $U_{A^*}$. They show that although (quantum) oracle access to $U_{A^*}$ is sufficient for quickly verifying $\ket{A}$, it still requires exponential time to clone $\ket{A}$ (or to create a state which has a non-negligible overlap with $\ket{A}\tensor\ket{A}$). They construct a public mini-scheme relative to an oracle that does not suffer from the attacks which will be described next, and is proved to be unconditionally secure. 

They also construct an explicit public quantum money mini-scheme (see Definition~\ref{def:quatnum_money}) without any oracles.
At the time, indistinguishability obfuscation was not well studied, and in particular, no candidate constructions were available. So, they invented an ad-hoc approach to obfuscate subspace membership in the following way.
The bank picks a random element $A \in S(n)$ as before. Next, it picks \nom{$p_1,\ldots,p_{\beta n}$}{Polynomials which obfuscate $\IA$} at random from the set of multi-variate polynomials with a fixed degree $d$ that vanish on $A$. This can be preformed efficiently. $\beta$ is chosen so that with exponentially high probability, the elements of $A$ are the only ones which vanish on all polynomials, i.e. $\{p_{i}\}_{i\in [\beta n]}$ can be used to compute $\IA$: 
\begin{equation}
  \label{eq:3}
  a\in A \ \iff \ p_{i}(a)=0 \ \   \forall i\in  [\beta n ]
\end{equation}

 The same process is done for $A^\perp$, with polynomials $q_1,\ldots,q_{\beta n}$. 

In the noise-free mini-scheme, these polynomials are published by the bank as the public key,
and provide a means to compute $\IAH$, using Eq.~\eqref{eq:3}, and its analogue for $A^{\perp}$.

In the noisy version, again, $\beta n$ polynomials are provided. This time, an $\epsilon$ fraction of them are ``decoy'' polynomials: these are degree $d$ polynomials that vanish on a random $A'\in S$ (where each polynomial has a different $A'$ associated with it). The other $(1- \epsilon) \beta n$ are as before (uniformly random from the set of all degree $d$ polynomials that vanish on $A$). The polynomials are provided at a random order. The parameters $\beta$ and $\epsilon$ are picked so that the vanishing polynomials can still be used to compute $\IAH$ using the following property:
\begin{equation}
\label{eq:5}
  a\in A \iff  |\{ p_{i}(a)=0\}_{i \in [\beta n]}| \geq (1-\epsilon)\beta n. 
\end{equation} 

Pena, Faug\`ere and Perret showed how to totally break (that is, they find a basis for $A$, and therefore reconstruct the secret key) the noise-free version, using Gr\"{o}bner's basis analysis~\cite{Pena15Algebraic}. 
\subsection{Attack on the Noisy Scheme}
\label{sec:attack-noisy-scheme}
In this section we report on an attack by Paul Christiano~\cite{Christiano2015private} which totally breaks the noisy version, by reducing the noisy version to the noise-free version.  Note that as of the time this attack was discovered, there was no attack known on the noisy scheme. 

Christiano's attack reduces the noisy version to the noise-free version by distinguishing between the polynomials that vanish on $A$ and the ``decoy'' polynomials. The attack requires access to one copy of the quantum money state, and is inherently quantum. Interestingly, the noise-free version attack, unlike ours, is purely classical and does not require a copy of the quantum money state.  

The attack is based on the notion of \emph{single copy tomography} which was introduced by Farhi et al.~\cite{farhi2010quantum}.\footnote{A similar result, based on different techniques and under weaker assumptions was discovered  earlier~\cite{aharonov1993meaning,aharonov1993measurement}, though the motivation and the formulation was different.} They showed that given a single copy of an unknown state $\ket{\psi}$, and a two outcome measurement $\{\ketbra{\psi}, I-\ketbra{\psi}\}$, which we call the protective measurement, it is possible to get a ``good'' estimate of $\bra{\psi} E_{j} \ket{\psi}$ efficiently for every efficiently implementable POVM\nom{ $\{E_{j}\}$}{POVM elements}, without destroying the state $\ket{\psi}$. For our purposes, we will be interested only in a 2 outcome measurement with an additive approximation $\frac{1}{10}$, and error probability $1/(\beta n)^{2}$, which, they show, requires only $O(\log(\beta n))$ uses of the protective measurement. We define, $E^{i}_{0}\ket{x}=p_{i}(x)\ket{x}$, and $E^{i}_{1}=I-E^{i}_{0}$ for every $i \in [\beta n]$.   Recall that since $p_{i}$ is a polynomial over $\mathbb{F}_{2}^{n}$, $p_{i}(x)\in \{0,1\}$, which implies that $\{E^{i}_{0},E_{1}^{i}\}$ is a POVM (in this case, a 2-outcome projective measurement), and furthermore, by standard techniques, since $p_{i}(x)$  can be calculated efficiently classically, this POVM is also efficiently implementable for every $i$. 
Since the "decoy" polynomials are fully random, $\bra{A} E^{i}_{0} \ket{A}\approx \frac{1}{2}$ with very high probability. On the other hand, for the polynomials $p_{i}$ which vanish on $A$, $\bra{A} E^{i}_{0} \ket{A} = 0$.

We use Farhi et al.'s single copy tomography to estimate $\bra{A} E^{i}_{0} \ket{A}$ for every $i\in [\beta n]$. 
By the union bound, with probability at least $1-\frac{1}{\beta n}$, the estimation for all the noisy polynomials is at least $0.3$, while for all the polynomials that vanish on $A$ the estimation is at most $\frac{1}{10}$. After sieving the vanishing polynomials from the ``decoys'', we can use Pena et al.'s noise-free attack.  

\begin{remark}
    In a subsequent work~\cite{PDF+19}, a different attack on the noisy-scheme, was found. The main advantages of the new attack is that it purely classical: the attacker only needs the public-key (cf. the attack above, which also requires one token), and only needs a classical computer (cf. the attack above, which needs a quantum computer for the single-copy tomography).
\end{remark}

\section{Memoryless Digital Signatures from Tokens}
\label{app:super-secure}

In this section, we show how to construct a standard (not memory dependent)
digital signature scheme from a tokenized signature scheme with some
extra security properties, which we call
super-security and unpredictability. Super-security, also known as strong unforgeability, is a stronger notion of security than unforgeability that was extensively studied also in the classical literature (see, e.g.,~\cite[Section 6.5.2]{goldreich04foundations}, and \cite[Definition 4.3]{KL14} in the context of MAC). Informally, in the classical setting, super-security guarantees that an adversary cannot generate a fresh signature even for a document which was signed by the signing oracle. In our setting, since there is no signing oracle, we define super-security as the inability of the forger which receives $\ell$ tokens to provide $\ell+1$ valid and distinct document \& signature pairs. So, for example, an adversary which receives one signing token and provides two different signatures for the same document is considered as successful forgery.  Towards providing a formal security game, we define the algorithm $\verify'_{k,pk}$, in which requirement \ref{it:verify_k_1} from the definition of $\verify_{k,pk}$ (see p.~\pageref{it:verify_k_1}) is changed to: all \emph{document \& signature pairs} are distinct, i.e. $(\alpha_{i},s_{i}) \neq (\alpha_{j},s_{j})$ for every $1 \leq i \neq j \leq k$.  

\begin{definition}[Super-security and One-time Super-security]
\label{def:super_security}
A TS scheme is super-secure if for every $\ell=\poly(\kappa)$ a QPT adversary cannot provide $\ell+1$ different signatures by using the public key and $\ell$ signing tokens:
\begin{equation}
	\Pr\left[\verify'_{\ell+1,pk}( \Adv(pk,\ket{\stamp_{1}} \tensor \ldots \ket{\stamp_{\ell}})) = 1 \right]\leq \negl(\kappa) 
	\label{eq:super_security}
\end{equation}
One-time super-security requires the above only for $\ell=1$.
\end{definition}

\begin{definition}[Unpredictability]
	A TS scheme is unpredictable if signing the same document twice, using two signing tokens, gives two different signatures except with negligible probability:
	\begin{equation}
	\Pr\left[\sign(\alpha,\ket{\stamp_{1}}) = \sign(\alpha,\ket{\stamp_{2}})\right] \leq \negl(\kappa)
	\end{equation}
	\label{def:unpredictability}
\end{definition}

We note that unpredictability for two documents implies unpredictability for polynomially many documents.

\begin{theorem}
	If $\ts$ is an unpredictable (see Definition~\ref{def:unpredictability}) and super secure (see Definition~\ref{def:super_security}) tokenized signature scheme then the scheme $\ds$ as defined in Section \ref{sec:properties-tokenized-signatures} is an unforgeable digital signature scheme (see Definition~\ref{def:digital_signature}).
	\label{thm:ts_implies_ds}
\end{theorem}

\begin{remark}
The proof of Theorem~\ref{thm:ts_implies_ds} could easily be adjusted to show that under the same assumptions, $\ts$ is unforgeable when the adversary with $\ell$ signing tokens is augmented with a signing oracle which uses the same $\sign$ procedure as in $\ds$. The adversary needs to provide $\ell+1$ valid signed documents as before, with the additional requirement that these documents were not signed by the signing oracle.
\label{rem:ts_with_oracle}
\end{remark}

\begin{remark}
Behera, Shinar and Sattath showed how to lift \emph{any}  secure TS scheme to a secure  digital signature scheme, using a different construction --- see~\cite[Appendix A.2.4]{BSS21}.\footnote{Even though their construction was in the context of tokens for MACs, the proof is valid also for tokens for digital signatures.} Therefore, the unpredictability and super-security assumptions in Theorem~\ref{thm:ts_implies_ds} are unnecessary if one is willing to settle for a different, slightly more complicated construction then $\ds$. 
\end{remark}

\begin{proof}[Proof of Theorem~\ref{thm:ts_implies_ds}]
	Assume towards a contradiction that $\ds$ is forgeable: there exists a QPT $\Adv$ for which,
	\begin{equation}
	\Pr\left[ (\alpha,sig) \gets \Adv^{\ds.\sign_{sk}}(pk),\ \ds.\verify_{pk}(\alpha,sig) = 1 \text{ and } \alpha \notin Q_{\Adv}^{\sign_{sk}}\right] \geq \nonnegl(\kappa).
	\label{eq:8}
	\end{equation}
	
	We define an adversary $\Adv'$ which receives as an input $\ell$ signing tokens (here $\ell = \poly(\kappa)$ is the number of oracle queries $\Adv^{\ds.\sign_{sk}}(1^{\kappa})$ applies). We will show that $\Adv'$ violates the super-security of $\ts$ and reach a constradiction. $\Adv'$ simulates $\Adv^{\ds.\sign}(1^{\kappa})$. Whenever $\Adv^{\ds.\sign}(1^{\kappa})$ makes an oracle query to sign the document $\alpha$, $\Adv'$ uses one of its signing tokens to sign that document. Suppose the documents that were given to the signing oracle as queries are $\alpha_{1},\ldots,\alpha_{\ell}$ and the outputs are $sig_{1},\ldots, sig_{\ell}$, and we denote the outputs $(\alpha,sig)$ of $\Adv^{\ds.\sign}$ by $(\alpha_{\ell+1},sig_{\ell+1})$. Overall, $\Adv'$ outputs $(\alpha_1,sig_1,\ldots,\alpha_{\ell+1},sig_{\ell+1})$.
	
	\begin{align}
	\Pr [ \verify'&_{\ell+1,pk}  (\Adv'(pk,\ket{\stamp_{1}}\tensor \ket{\stamp_{\ell}}))=1 ]  \\
	&= \Pr\left[ \verify'_{\ell+1,pk}(\alpha_{1},sig_{1},\ldots,\alpha_{\ell+1},sig_{\ell+1})=1\right] \\
	& =  \Pr\left[ \verify_{pk}(\alpha_{\ell+1},sig_{\ell+1})=1 \text{ and }  (\alpha_{i},sig_{i}) \neq (\alpha_{j},sig_{j})\ \forall i\neq j \leq \ell+1 \right] \\
	& \geq  \Pr\left[ \verify_{pk}(\alpha_{\ell+1},sig_{\ell+1})=1 \text{ and } \forall i \leq \ell,\ (\alpha_{i},sig_{i}) \neq (\alpha_{\ell+1},sig_{\ell + 1}) \right] -\negl(\kappa)\\
	& \geq  \Pr\left[ \verify_{pk}(\alpha_{\ell+1},sig_{\ell+1})=1 \text{ and } \forall i \leq \ell,\ \alpha_{i} \neq \alpha_{\ell+1} \right] -\negl(\kappa)\\
	& =  \Pr\left[(\alpha,sig)\gets \Adv^{\ds.\sign_{sk}}(pk):\ \ds.\verify_{pk}(\alpha,sig)=1  \text{ and } \alpha\notin Q_{\Adv}^{\sign_{sk}} \right] - \negl(\kappa)\\
	&\geq \nonnegl(\kappa)
	\label{eq:9}
	\end{align}
	The second step is justified by our assumption that $(\alpha_{1},sig_{1}), \ldots, (\alpha_{\ell},sig_{\ell})$ are the outputs of the signing oracle and therefore pass the verification test by the correctness property. The unpredictability is used in the third step, and Eq.~\eqref{eq:8} for the last step. 
	To conclude, Eq.~\eqref{eq:9} shows that $\ts$ is not super-secure, contradicting our assumption. 
\end{proof}

By removing the second inequality in Eq.~\eqref{eq:9} and adapting the following step, the proof above can be strengthened to show that $\ds$ is a super-secure digital signature scheme -- see~\cite[Section 6.5.2]{goldreich04foundations}.

\subsection{Proving Super-Security and Unpredictability}
\label{app:proving_super_security_and_unprdictability}
We show that our construction of tokenized signatures
is super-secure and unpredictable (under the same obfuscation
assumption as before).

We first claim that all our reductions maintain super-security (Definition~\ref{def:super_security}) and unpredictability (Definition~\ref{def:unpredictability}). It is easy to see that
if a one-bit tokenized scheme gives different signatures each time
a new pair of keys is generated (except with negligible probability),
then the resulting full tokenized signature scheme will be unpredictable.
Showing that a one-time super-secure tokenized signature scheme
gives rise to a full super-secure tokenized signature scheme is
a bit more involved, but directly follows the proof of Theorem~\ref{thm:standard_construction_unforgability}.

Next, we show that the oracle scheme is one-time super-secure.

\begin{theorem}\label{thm:oracle-super-secure}
	Let $A$ be a uniformly random subspace from $S(n)$,
	and let $\epsilon>0$ be such that $1/\epsilon=o\left(2^{n/2}\right)$.
	Given one copy of $\ket{A}$ and a membership oracle for $A$,
	a counterfeiter needs $\Omega(\sqrt{\epsilon} 2^{n/4})$ queries
	to output a pair $(a,b)\in (A\setminus\{0\})\times (A\setminus\{0\})$
	satisfying $a\neq b$ with probability at least $\epsilon$.
	The same number of queries is also required to output a pair
	$(a,b)\in (A^\perp\setminus\{0\})\times (A^\perp\setminus\{0\})$
	satisfying $a\neq b$ with probability at least $\epsilon$.
\end{theorem}

\begin{proof}
First, note that finding such a pair of elements in $A$
and finding such a pair in $A^\perp$ are essentially the same task
(up to renaming $A$ and $A^\perp$); it suffices to prove the result
for a pair in $A$, and the result for $A^\perp$ will follow by symmetry.

Our proof will be extremely similar to the proof of
Theorem~\ref{thm:lower-bound-average-case}.
In fact, if we simply redefine $\Lambda(A)$ to be the set
$\{(a,b)\in A^2:a\neq b, a\neq 0, b\neq 0\}$,
the whole proof goes through except for Lemma~\ref{lem:probability_a_in_B_upperbound}.
We therefore prove an analogue of this lemma with the redefined
$\Lambda(A)$.

Fix $A\in S(n)$ and $(a,b)\in \Lambda(A)$,
so $a\neq b$, $a,b\in A$, and $a,b\neq 0$.
Picking $B\sim\mathcal{R}_A$ is picking a vector
space whose intersection with $A$ has dimension $n/2-1$.
One way of picking such a vector space uniformly
at random is by picking a random basis for $A$,
then discarding one of the vectors in the basis
and adding a vector outside of $A$. This gives
a new set of $n/2$ independent vectors, which
we take to be a basis for $B$.

We are interested in the probability that $(a,b)\in \Lambda(B)$.
This is simply the probability that $a,b\in B$,
which is equal to the probability that $a$ and $b$
are both in the span of a randomly chosen set of
$n/2-1$ independent vectors in $A$.

Now, the number of subspaces of $A$ of dimension $n/2-1$
is $G(n/2,n/2-1)$, and the number of such
subspaces that contain $a$ and $b$ is simply
$G(n/2-2,n/2-3)$. The probability that a random subspace
contains $a$ and $b$ is therefore
\[\frac{G(n/2-2,n/2-3)}{G(n/2,n/2-1)},\]
which simplifies to
\[\frac{2^{n/2-2}-1}{2^{n/2}-1}
=\frac{1}{4}\left(1-\frac{3}{2^{n/2}-1}\right)
\leq\frac{1}{4}.\]
\end{proof}

Finally, we show that the construction from a VBB obfuscator
with dependent auxiliary inputs is one-time super-secure, which
implies that under our strong obfuscation assumption our scheme
is unforgeable and super-secure (and thus can be used as a digital
signature scheme).

\begin{theorem}
	If $\mathcal O$ is a VBB obfuscator with dependent auxiliary input for the class $\mathfrak S$, then the scheme $\ot1$ described in Algorithm~\ref{alg:ot1} is one-time super-secure.
\end{theorem}
\begin{proof}
	The proof is essentially identical to the proof
	that the scheme is unforgeable, except we use
	$\verify'$ instead of $\verify$. Fix $A$. Suppose the distinguisher 
	runs the (classical) algorithm $\verify'_{2, pk(A)}$,
	and the auxiliary state is $\ket{A}$. By the VBB property,
	\[ |\Pr_{\Adv, \mathcal O}[\verify'_{2,pk(A)}(\Adv(\mathcal O(\IAH),\ket{A}))=1]- \Pr_{\Sim}[\verify'_{2,pk(A)}(\Sim^{\IAH}(1^{|\IAH|},\ket{A}))=1] |\leq \negl(\kappa). \] 
	Here, we were explicit over what the probability is.
	
	The above holds for every $A \in S$. We now take the average over all $A \in S$:
	\[ |\Pr_{A \getsr S, \Adv, \mathcal O}[\verify'_{2,pk(A)}(\Adv(\mathcal O(\IAH),\ket{A}))=1]- \Pr_{A \in_{R} S,\Sim}[\verify'_{2,pk(A)}(\Sim^{\IAH}(1^{|\IAH|},\ket{A}))=1] |\leq \negl(\kappa) \] 
	By Theorem~\ref{thm:oracle-super-secure}, the second term of the above equation can be bounded. For every QPT $\Sim$,
	\begin{equation}
	\Pr_{A \getsr S,\Sim}[\verify'_{2,pk(A)}(\Sim^{\IAH}(1^{|\IAH|},\ket{A}))=1]=\frac{\poly(\kappa)}{2^{n/2}}=\negl(\kappa)
	\end{equation}
	Combining the above two equations, we conclude that one-time super-security (see Definition~\ref{def:super_security}) holds:
	\begin{equation}
	\Pr_{A\getsr S, \Adv, \mathcal O}[\verify'_{2,pk(A)}(\Adv(\mathcal O(\IAH),\ket{A}))=1]\leq \negl(\kappa).
	\end{equation}
\end{proof}

\section{Tokens for Message Authentication Codes}
\label{sec:MAC}
The goal of this section is to construct a \emph{private} tokenized signature scheme. The main advantage is that here, unlike the public variant, there is no need for obfuscation, and our construction can be instantiated based on standard generic assumptions.

\subsection{Definitions}
Our construction will be based on post-quantum CMA-secure MACs (the private analog of a digital signature), and post-quantum CPA-secure private-key encryption. 


\begin{definition}[Message Authentication Code]\label{def:MAC}
	A message authentication code (MAC) consists of 3 PPT algorithms $\keygen,\ \sign$ and $\verify$ satisfying:
	\begin{enumerate}
    	\item $\keygen(1^\kappa)$ outputs a key $k$, where $\kappa$ is the security parameter.
		\item When a document is signed using the signing key, the signature
		is accepted by the verification algorithm using the verification key.
        That is, for every $\alpha\in \{0,1\}^{*}$,
		\begin{equation}
		\Pr\left[\verify_{k}\left(\alpha,\sign_{k}\left(\alpha\right) \right)= 1 \right] = 1
		\label{eq:mac_correctness}
		\end{equation}
		\item The scheme is secure against quantum existential forgery under chosen message
		attacks; that is, a quantum adversary with the capability of adaptively requesting
		\emph{classical} documents to be signed by a signing oracle
		cannot generate a signature for a document that was not queried:
		\begin{equation}
		\Pr\left[
		(\alpha,sig) \gets \Adv^{\sign_{k}},\ \verify_{k}(\alpha,sig) = 1 \text{ and } \alpha \notin Q_{\Adv}^{\sign_{k}}\right] \leq \negl(\kappa),
		\end{equation} \label{eq:unforgeability_private_ts} 
		where $\Adv^{\sign_{k}}$ is a QPT algorithm with access to a classical signing
		oracle, and $Q_{\Adv}^{\sign_{k}}$ is the set of documents the signing oracle has signed.
	\end{enumerate}
\end{definition}

\begin{definition}[{Post-quantum CPA-secure Private Key Encryption, adapted from~\cite[Section 3.4.2]{KL14}}]
\label{def:encryption}
A post-quantum secure secure against chosen plaintext attacks (or CPA-secure) private key encryption scheme consists of $3$ PPT algorithms,
$\keygen$, $\encrypt$, and $\decrypt$, with the following properties:
\begin{enumerate}
    \item On input $1^\kappa$ where $\kappa$ is the security parameter, $\keygen$ outputs a key $e$.
	\item $\encrypt_{e}$ and $\decrypt_{e}$ are maps from $\{0,1\}^*$ to
    $\{0,1\}^*$ such that for all $\alpha\in\{0,1\}^*$,
    \[\Pr\left[\decrypt_{e}(\encrypt_{e}(\alpha))=\alpha\right]\geq 1-\negl(\kappa).\]
    \item For every QPT adversary $\Adv$,
    \begin{equation}
        \Pr[\Adv \text{ wins the CPA indistinguishability experiment}]\leq \frac{1}{2}+\negl(\kappa).
    \end{equation}
\end{enumerate}
The CPA (chosen-plaintext attack) indistinguishability experiment is defined as follows:
\begin{enumerate}
    \item A key $e\gets \keygen(1^\kappa)$ is generated.
    \item $\Adv$ is given input $1^\kappa$ and \emph{classical} oracle access to $\encrypt_{e}$, and outputs a pair of massages $m_0,m_1$ of the same length.
    \item A uniform bit $b\in \{0,1\}$ is chosen, and then a ciphertext $c\gets \encrypt_k(m_b)$ is computed and given to $\Adv$.
    \item The adversary $\Adv$ continues to have oracle access to $\encrypt_k$ and outputs a bit $b'$.
    \item The adversary wins iff $b=b'$.
\end{enumerate}
\end{definition}

We define a private (length-restricted)  tokenized signature scheme
in analogy to a public (length-restricted) public tokenized digital signature scheme.
The only difference occurs in the symmetric key, and the definition of unforgeability.

\begin{definition}[Private tokenized signature scheme (private TS)]
	\label{def:private_TS}
	A private TS scheme consists of 4 QPT algorithms, \keygen, \tokengen, \sign, and \verify, with the following properties:
	\begin{enumerate}
		\item On input $1^{\kappa}$ where $\kappa$ is the security parameter, $\keygen$ outputs a classical  key $k$.
		\item $\tokengen_{k}$ generates a signing token $\ket{\stamp}$. We emphasize that if $\tokengen_{k}$ is called $\ell$ times it may (and in our construction, would) output different states $\ket{\stamp_{1}},\ldots,\ket{\stamp_{\ell}}$.
		\item\label{it:correctness} For every document $\alpha \in \{0,1\}^{*}$, 
		\[ \Pr\left[\ket{\stamp}\gets \tokengen_{k},\verify_{k}(\alpha,\sign(\alpha,\ket{\stamp}))=1 \right]=1.\]
		In an imperfect scheme, the above requirement is relaxed:
		\begin{equation}
		\Pr\left[\ket{\stamp}\gets \tokengen_{k},\verify_{k}(\alpha,\sign(\alpha,\ket{\stamp}))=1 \right]\geq 1 - \negl(\kappa) .
		\label{eq:imperfect_private_TS} 
		\end{equation}
	\end{enumerate}
\end{definition}

\begin{definition}[Length restricted private TS]\label{def:length_restricted_private_TS}
A private TS is $r$-restricted if Item~\ref{it:correctness} holds only for $\alpha \in \{0,1\}^{r}$.
\end{definition}

To define unforgeability, we introduce the algorithm $\verify_{\ell,k}$. This algorithm takes as an input $\ell$ pairs $(\alpha_{1},s_{1}),\ldots,(\alpha_{\ell},s_{\ell})$ and accepts if and only if (i) all the documents are distinct, i.e. $\alpha_{i}\neq \alpha_{j}$ for every $1 \leq i \neq j \leq \ell$, and (ii) all the pairs pass the verification test $\verify_{k}(\alpha_{i},s_{i})$.

\begin{definition}[Unforgeability and one-time unforgeability]\label{def:unforgeability_private_TS}
A private TS scheme is unforgeable if for every $\ell=\poly(\kappa)$ a QPT adversary cannot sign $\ell+1$ different documents by using $\ell$ signing tokens:
	\begin{equation}
	\Pr\left[\verify_{\ell+1,k}( \Adv(\ket{\stamp_{1}} \tensor \ldots \ket{\stamp_{\ell}})) = 1 \right]\leq \negl(\kappa) 
	\label{eq:unforgeability_private_TS}
	\end{equation}
	One-time unforgeability requires the above only for $\ell=1$.
\end{definition}

\begin{remark}
Ref.~\cite{BSS21} considered stronger definitions for unforgeability, where the adversary has a classical or quantum oracle access to $\verify_k$. They show that by using a slightly stronger primitive (namely, they use a strong authenticated encryption scheme), Lemma~\ref{lem:ot_to_full_TMAC} could be strengthened to handle this stronger security model, though their construction is slightly more complex.
\end{remark}

\subsection{From a 1-bit Onetime Scheme to a Full scheme}
Consider the 1-bit length restricted private TS scheme given in Algorithm~\ref{alg:mac_ot1}. The (imperfect) correctness requirement hold for the same reasons as the public scheme. 
Theorem~\ref{thm:lower-bound-average-case} shows that it is one-time unforgeable, even against computationally unbounded adversaries: the theorem shows that an adversary which is given $\ket{A}$ needs exponentially many queries to the oracle are needed in order to find a non-trivial element in $A$ and a non-trivial element in $A^\perp$ with non-negligible probability. But in our setting, the adversary have no oracle access, and therefore clearly his success probability is negligible. 

\begin{algorithm}
  \caption{$\ot1$ private tokenized signature scheme.}
   
    \begin{algorithmic}[1] 
        \Procedure{$\keygen$}{$1^{\kappa}$}
            \State  Set $n=f(\kappa)$ \Comment{$f$ can be any super-logarithmic function.} 
            \State $A\getsr S(n)$
            \State \textbf{return} $sk \equiv  \langle A\rangle $ \Comment{$\langle A \rangle$ is a (classical) basis for $A$. }
        \EndProcedure

        \Procedure{$\tokengen$}{$\langle A \rangle$}
            \State $\ket{\stamp} \equiv \ket{A}$ \Comment{Given a basis for $A$, the state $\ket{A}$ can be generated efficiently~\cite{aaronson2012quantum}.}
            \State \textbf{return} $\ket{\stamp}$
        \EndProcedure

       \Procedure{$\sign$}{$\alpha \in \{0,1\},\ket{A}$}
            \State Apply $H^{\otimes n}$ iff $\alpha=1$ to the state $\ket{A}$. 
 \Comment{$H^{\tensor n}\ket{A}=\ket{A^{\perp}}$}
            \State Measure the resulting state  in the standard basis. Set $sig$ to be the outcome. 
            \If{$sig=0 \ldots 0$}
                \State \textbf{return} failed
            \Else
                \State \textbf{return} $sig$
            \EndIf
        \EndProcedure

        \Procedure{$\verify_{sk}$}{$\alpha \in \{0,1\}, sig$}            
            \If{$sig=0^n$}
                \State \textbf{return} $0$
            \EndIf
            \State \textbf{return} $\IAH(\alpha,sig)$.
        \EndProcedure
    \end{algorithmic}
\label{alg:mac_ot1}
\end{algorithm}

The goal of the rest of this section is to extend a one-time $1$-bit length restricted private TS scheme,
into a full private TS scheme. This is done by applying the analogous transformations in the public setting. The transformation from 1-restricted to r-restricted (see Lemma~\ref{le:mac_1_bit_to_r_bits}) and from $r$-restricted to unrestricted (see Lemma~\ref{le:mac_r_bits_to_unrestricted}) is the same, except the cosmetic adaptations needed. 
The transformation from a one-time scheme to a full scheme (see Lemma~\ref{lem:ot_to_full_TMAC}) is slightly different. Recall that in the public setting, the public key scheme contained a (digitally signed) obfuscated program that tests membership in $A$ and in $A^\perp$.  Here, we want to avoid that, yet giving someone holding the secret key to evaluate these membership functions. So, instead
of providing this program in an obfuscated form, we encrypt the basis of $A$ using a post-quantum CPA-secure private-key encryption. 

\begin{lemma}
	A $1$-restricted one-time private TS $\ot1$ can be transformed into an
	$r$-restricted private TS $\otr$, such
	that if $\ot1$ is unforgeable, so is $\otr$.
	Here $r$ is an integer which is at most polynomial in the security
	parameter $\kappa$.
	\label{le:mac_1_bit_to_r_bits}
\end{lemma}

\begin{proof}
As with public tokenized signature schemes, we construct $\otr$ from $\ot1$
by repeating it $r$ times; that is, $\keygen$ for $\otr$ will run $\keygen$
for $\ot1$ $r$ times, and use the tuple of signing keys as its signing key
and the tuple of verification keys as its verification key.
Signing a document $\alpha\in\{0,1\}^r$ works by simply signing
each bit of $\alpha$ separately. Soundness follows from the soundness
of the $\ot1$ scheme; that is, the verification of a valid signature for $\alpha$
fails to accept if and only if one of the verifications for the $r$ copies
of the $\ot1$ scheme fails to accept. By the union bound, this is at most
$r$ times the probability that $\ot1$ fails to verify a valid signature;
this is $0$ in a perfect scheme and $r\cdot\negl(\kappa)$ in an
imperfect scheme.

We show unforgeability. Suppose there was an adversary $\Adv$
that signed two different documents of length $r$
with non-negligible probability $f(\kappa)$.
Consider a new adversary $\Adv'$ that attempts to sign two different documents
of length $1$. $\Adv'$ takes as input one signing token. It then runs
$\keygen$ and $\tokengen$ $r-1$ times, to get $r-1$ additional signing
tokens. It shuffles these together, and uses the $\Adv$ algorithm
to sign two different documents (for $\otr$) with non-negligible
probability $f(\kappa)$. Since the documents are different, they must differ
on some bit; there is a $1/r$ chance that this bit is the same bit as the
``real'' input token, rather than the artificially generated ones. In that case,
$\Adv'$ successfully created valid signatures for two different documents
relative to $\ot1$. This happens with probability $f(\kappa)/r$, which
is non-negligible, giving a contradiction.
\end{proof}

\begin{lemma}
	An $r$-restricted one-time private TS $\otr$ can be transformed into a
	one-time unrestricted private TS $\ot$ with the help
	of a hash function, such that if $\otr$ is unforgeable and the hash
	function is collision-resistant, then $\ot$ is unforgeable.
	\label{le:mac_r_bits_to_unrestricted}

\end{lemma}

\begin{proof}
	The $\ot$ scheme works by the hash-and-sign paradigm: we first
	hash a document down to length $r$, and then sign the hash.
	Here $r$ needs to be a function of the security parameter $\kappa$,
	say $r=\kappa$. The soundness of the scheme is clear.
	The unforgeability also follows easily: if an adversary
	produces valid signatures for two documents, then either the two
	documents have the same hash value (which contradicts
	the collision-resistance of the hash function), or else
	the two documents have different hash values, which contradicts
	the unforgeability of the $\otr$ scheme.
\end{proof}

\begin{lemma}
	A one-time private TS $\ot$ can be transformed into a full-blown private TS
	$\tm$ with the help of a quantum secure MAC and a post-quantum CPA-secure private-key encryption scheme.
	\label{lem:ot_to_full_TMAC}
\end{lemma}
\begin{proof}
The scheme $\tm$ will work as follows. The algorithm $\keygen$ will generate
a key $k$ for the classical MAC, as well as a key $e$ for the
classical encryption scheme, using the security parameter $\kappa$. The pair $(k,e)$ will be the key
of $\tm$.

The algorithm $\tokengen_{k,e}$ will then use $\ot$ to generate
a key $k_1$ and a signing token $\ket{\stamp_1}$;
it will then encrypt $k_1$ using the encryption key $e$, and sign
the resulting encrypted message using the MAC and $k$.
The final signing token will be $\ket{\stamp_1}$ appended with
the encrypted key $\encrypt_e(k_1)$, and the its signature $MAC.\sign_k(\encrypt_e(k_1))$.

The new signing algorithm $\sign$ applies the signing procedure of $\ot$,
and appends to this signature the signed encrypted verification key that came
as part of the token.

The algorithm $\verify$ uses $k$ to verify the signature of the encrypted key,
uses $e$ to decrypt the key $k_1$,
and uses $k_1$ to verify the signature according to the $\ot$ protocol.

It's clear that the verification procedure accepts the output of the
signing procedure (assuming this holds for $\ot$). We now show
that if $\ot$ is unforgeable, then so is $\tm$.

Suppose by contradiction that there was an adversary $\Adv$
that used $\ell$ signing tokens and produced $\ell+1$ document \& signature pairs
$a_1,a_2,\dots,a_{\ell+1}$
that are distinct and accepted with non-negligible probability $f(\kappa)$
by the verification algorithm. Each $a_i$ must have the form
$a_i=(\alpha_i,s_i,q_i,t_i)$ where $\alpha_i$ are the distinct documents
that were signed, $t_i$ represents a MAC signature (also known as tag) for $q_i$, $q_i$ represents the encrypted verification key, and $s_i$ represents
the $\ot$ tokenized signature for $\alpha_i$. If the verification algorithm accepts the
$(a_1,a_2,\dots,a_{\ell+1})$ tuple, we know that the $\alpha_i$ documents
are all distinct, and that the verification algorithm accepts each $a_i$.
In turn, this tells us that $t_i$ is a valid MAC signature for $q_i$, and that $\decrypt_{e}(q_i)$ outputs a string $p_i$
with the property that $\verify_{p_i}(\alpha_i,s_i)$ accepts.

The fact that $t_i$ is a valid signature for $q_i$ means that $q_i$ was
signed using the classical MAC with non-negligible probability
(otherwise, the MAC is broken by $\Adv$). Since $\Adv$ does not have
access to the signing key or verification key for the MAC, by the MAC security, the $(q_i,t_i)$
must be signatures that were appended to the signing tokens. This means there
are only at most $\ell$ unique values for the $q_i$, so by the pigeonhole
principle, some two are equal; without loss of generality, say $q_1=q_2$.

Let $q=q_1=q_2$, and let $p$ be the decryption of $q$ using $e$.
Since the signature $t_1$ of $q$ is valid, we know that $q$ is really
one of the messages signed by the token-printing authority, so $p$
is a key for one of the copies of the one-time protocol.
The adversary received one token, $\ket{\stamp}$, that corresponds
to the key $p$. Since $a_1$ and $a_2$ are accepted, we know that
$\verify_p(\alpha_1,s_1)$ and $\verify_p(\alpha_2,s_2)$
both accept with non-negligible probability;
that is, the adversary produced two valid signatures for the one-time
protocol. This is almost a contradiction, but not quite:
the adversary did produce two valid signatures using one token for
the one-time protocol, but it had access to a bit more information --
an encryption of the verification key for that one-time protocol.

In other words, using the hypothesized adversary for the private TS $\tm$,
we've produced an adversary for $\ot$ that takes in the token and
an encryption of the key, and produces two valid signatures
(for different documents). To obtain a contradiction, we must show
that the adversary must either break the encryption, or else break
the unforgeability of $\ot$.

We do this by considering what happens when we feed this adversary
a valid token but a bogus encryption (e.g., the encryption of a random
string rather than the verification key). There are two cases:
either the adversary fails to produce two valid signatures with non-negligible
probability, or else it succeeds. If it succeeds, then it breaks the $\ot$
scheme, which is a contradiction. Thus, suppose the algorithm fails.
In this case, by a standard argument, we can construct an adversary that has the ability to distinguish the real encryption of the verification key $p$ from the encryption of a random
string, violating the post-quantum CPA\footnote{Note that this new adversary needs chosen plaintexts, since the adversary also needs to encrypt the other $\ot$-TS keys.}-security of the private-key encryption scheme. Therefore, we reached the contradiction that no such adversary $\Adv$ could exist to begin with, and hence the private TS scheme $\tm$ is indeed unforgeable.
\end{proof}

We conclude that by combining the results in this section, we can construct a
full unforgeable private TS scheme, based on the existence of a post-quantum CMA-secure MAC scheme, a post-quantum
CPA-secure private-key encryption scheme, and a post-quantum collision-resistant
hash function.
Post-quantum one-way functions imply post-quantum CMA-secure MAC (and in fact, even CMA-secure digital signature) and  post-quantum private-key encryption~\cite{BZ13}. Additionally, almost trivially, a collision-resistant hashing scheme implies a one-way functions. Therefore, all three of these cryptographic
primitives follow from the existence of a collision-resistant hash function,
so this is the only assumption in our construction.


Also all the other  results that apply for public tokenized signatures have analogous statements for private tokenized signature schemes.
By arguments similar to those in Appendix~\ref{app:super-secure},
our $1$-bit private TS scheme is information-theoretically super-secure,
and our full TS scheme is super-secure as well (using a collision-resistant
hash function). Also, again by analogous arguments to
Appendix~\ref{app:super-secure}, a super-secure private TS can be used as a regular MAC. Recall that the existence of a MAC follows from the existence of a
one way function~\cite{goldreich04foundations}. What we've shown here is that using a slightly stronger
assumption -- specifically, a collision-resistant hash function
instead of a one way function --
we can construct a super-secure tokenized TS scheme, which is strictly
stronger than a regular MAC.

Essentially by the same proof of their public analogs, it can be shown that every private tokenized scheme is also revocable (the analog of Theorem~\ref{thm:revocability}), that every private testable tokenized signature scheme is a private quantum money scheme (the analog of Theorem~\ref{thm:testable_ts_is_qm}), and that the reductions we use to strengthen a 1-bit one-time private tokenized signature scheme to a full  private tokenized signature scheme respect testability (analogous to Theorem~\ref{thm:standard_construction_testable}) and everlasting revocability (Theorem~\ref{thm:standard_construction_everlasting}). 
\end{document}